\documentclass[10pt,conference]{IEEEtran}
\ifCLASSINFOpdf
\else
\fi
\usepackage{url}
\usepackage{color}
\usepackage{subfigure}
\usepackage{multirow}
\usepackage{bbding}
\usepackage{balance}
\let\labelindent\relax
\usepackage{enumitem,kantlipsum}
\usepackage[normalem]{ulem} 
\usepackage{graphicx}

\newif\ifdebugdoc\debugdocfalse
\ifdebugdoc
\newcommand{\add}[1]{\textcolor{red}{#1}}
\newcommand{\del}[1]{\textcolor{blue}{\sout{#1}}}

\else
\newcommand{\scott}[1]{}
\newcommand{\add}[1]{#1}
\newcommand{\del}[1]{}
\fi

\begin{document}
%
\title{The Insecurity of Home Digital Voice Assistants \\ --  Amazon Alexa as a Case Study}



%
\author{\IEEEauthorblockN{Xinyu Lei\IEEEauthorrefmark{1},
Guan-Hua Tu\IEEEauthorrefmark{1},
Alex X. Liu\IEEEauthorrefmark{1},
Kamran Ali\IEEEauthorrefmark{1},
Chi-Yu Li\IEEEauthorrefmark{2},
Tian Xie\IEEEauthorrefmark{1}}
\IEEEauthorblockA{\IEEEauthorrefmark{1} Michigan State University,
East Lansing, MI, USA\\ Email: leixinyu@msu.edu, ghtu@msu.edu, alexliu@cse.msu.edu, alikamr3@cse.msu.edu, xietian1@msu.edu}
\IEEEauthorblockA{\IEEEauthorrefmark{2} National Chiao Tung University, Hsinchu City, Taiwan \\Email:  chiyuli@cs.nctu.edu.tw}}


\IEEEoverridecommandlockouts
\makeatletter\def\@IEEEpubidpullup{5\baselineskip}\makeatother
\IEEEpubid{\parbox{\columnwidth}{$\bullet$ \textbf{This paper is officially published at CNS 2018 \cite{lei2018insecurity}.}
}
\hspace{\columnsep}\makebox[\columnwidth]{}}

\maketitle

\begin{abstract}

Home Digital Voice Assistants (HDVAs) are getting popular in recent years. Users can control smart devices and get living assistance through those HDVAs (e.g., Amazon Alexa, Google Home) using voice.
In this work, we study the insecurity of HDVA service by using Amazon Alexa as a case study. We disclose three security vulnerabilities which root in the insecure access control of Alexa services. We then exploit them to devise two proof-of-concept attacks, home burglary and fake order, where the adversary can remotely command the victim's Alexa device to open a door or place an order from Amazon.com.
The insecure access control is that the Alexa device not only relies on a single-factor authentication but also takes voice commands even if no people are around. We thus argue that HDVAs should have another authentication factor, a physical presence based access control; that is, they can accept voice commands only when any person is detected nearby.
To this end, we devise a Virtual Security Button (\texttt{VSButton}), which leverages the WiFi technology to detect indoor human motions. Once any indoor human motion is detected, the Alexa device is enabled to accept voice commands.
Our evaluation results show that it can effectively differentiate indoor motions from the cases of no motion and outdoor motions in both the laboratory and real world settings.
\end{abstract}



\section{Introduction}
\label{sect:intro}


In recent years, more and more home digital voice assistant (HDVA) devices are deployed at home. Its number is forecasted to grow thirteen-fold from 2015 (1.1 million) to 2020 (15.1 million), a compound annual growth rate of 54.74\%~\cite{StrategyDVA2017}. Thanks to the continuous efforts of the leading HDVA device manufacturers (e.g., Amazon and Google) and the third party voice service developers (e.g., CapitalOne, Dominos, Honeywell), users can do a great number of things using voice commands. They include playing music, ordering pizzas, shopping online, scheduling an appointment, checking weather, making a payment, controlling smart devices (e.g., garage doors, plug, thermostats), to name a few. To provide users with usage convenience, most of HDVA devices (e.g., Amazon Echo, \add{Google Home}) adopt an always-listening mechanism which takes voice commands all the time. Specifically, users are not required to press or hold a physical button on HDVA devices before speaking commands. 

However, such great convenience may expose users to security threats due to the openness nature of voice channels. Both owners and adversaries can speak commands to HDVA devices. 
At this point, the natural question is: \textit{Do these commercial off-the-shelf (COTS) HDVAs employ necessary security mechanisms to authenticate users and protect users from acoustic attacks?}

Unfortunately, our study on Amazon Alexa devices yields a negative answer. We identify three security vulnerabilities from them and devise two proof-of-concept attacks. The victims may suffer from home security breach and fake order attacks.
All the parties including the Alexa service provider (i.e., Amazon), Alexa devices, and the third party voice service developers, shall take the blame. The Alexa service employs only a single-factor authentication method based on a password-like voice word (e.g., ``Alexa''). For any person/machine who speaks the correct authentication word ahead of his/her voice command, the command can be accepted by the Alexa device. The device accepts voice commands no matter whether any persons are around. It works for all the sounds whose sound pressure level (SPL) is higher than 60dB. For the third party voice services, especially for Alexa-enabled smart device vendors, no access control is deployed at smart devices, since they assume all the voice commands from the Alexa service are benign.

Note that the reasons why we consider Amazon Alexa devices as a case study to explore the insecurity of HDVAs are as follows. First, they are the most popular and the best-selling flagship HDVA devices. According to a recent report, Alexa devices have been sold for over 5 million within two years since launch \cite{Sales}. Second, the Alexa service offers users more than 10,000 skills (Alexa voice services)~\cite{10000Skills} which are far more than its competitors. We believe that the security vulnerabilities discovered from the Alexa service/devices can be representative and the proposed remedy can be further generalized to other HDVA devices (e.g., Google Home).

At first glance, the remedy seems to be straightforward. HDVA devices shall authenticate users by their voice biometrics before taking voice commands. However, on the second thought, it may not be an easy task due to two reasons. First, users' voices may vary with their ages, illness, or tiredness. Second, human voice is vulnerable to replay attacks. Some of the prior works are proposed to deploy wearable devices for user authentication. For example, \cite{feng2017continuous} develops a proprietary wearable device which collects the skin vibration signals of users. The collected vibration signals are then continuously matched with the voice signals received by HDVA devices. However, users may be reluctant to wear this device at home all the time. Another solution is to force users to press a physical button to explicitly active Alexa devices before using it. Therefore, the main challenge for the remedy is \emph{how to effectively secure HDVAs without scarifying user convenience or introducing extra deployment cost?}


After a careful study on the security vulnerabilities discovered, we observe that acoustic attacks are mainly launched while victims are not at home; otherwise, these acoustic attacks will be heard by the victims. If HDVA devices stop taking voice commands when there are no surrounding users, the adversaries' fraudulent voice commands will not be accepted. To this end, we propose to deploy a Virtual Security Button (\texttt{VSButton}) on the HDVA devices. The \texttt{VSButton} system leverages COTS WiFi infrastructure deployed at home to detect tiny indoor human motions (e.g., waving a hand). Once indoor motions are detected, \texttt{VSButton} enables the microphone of HDVA devices to take voice commands for a period of time (e.g., 1 minute). Our experimental results show that \texttt{VSButton} can accurately differentiate indoor motions from the cases of no motions and outdoor motions. In our 100-minute experiment with the participation of six people, we do not observe any false alarms (i.e., the tested HDVA device is enabled by outdoor motions.).

In summary, we study the insecurity of HDVAs by systematically exploring all the parties involved: HDVA devices (e.g., Amazon Echo Dot), HDVA voice service providers (e.g., Amazon) and third party voice service developers (e.g., Garageio). This paper makes three major contributions.
\begin{enumerate}
\item We unveil three vulnerabilities of HDVAs towards acoustic attacks. All parties involved may share the blame.
 HDVA voice service providers merely employs weak single-factor authentication for users; HDVA voice services do not have physical presence based access control; the HDVA third party voice service developers do not enforce security policies on their connected devices.
\item We devise two proof-of-concept attacks (i.e., home burglary and fake order) based on the identified vulnerabilities. We access their impact in practice (in a controlled manner).

\item We design and develop a novel security mechanism (Virtual Security Button) which improves the security of HDVA voice services without scarifying user convenience. Our evaluation results show that small indoor human motions (e.g., wave a hand) can activate the HDVAs, whereas big outdoor motions (e.g, jump) cannot. The remedy proposed can be further applied to all home digital voice assistants supporting WiFi with proper software upgrade.
\end{enumerate}

The rest of this paper is structured as follows.
$\S$\ref{sect:prelim-threat-model} introduces Alexa service model and the adopted threat model.
We present three security vulnerabilities of home digital voice assistants, and sketch two proof-of-concept attacks in $\S$\ref{sect:insecure-voice-auth}.
We then propose the remedy, \texttt{VSButton}, in $\S$\ref{sect:sol}, evaluate its performance in real-world settings in $\S$\ref{sect:eval} and discuss several remaining issues in $\S$\ref{sect:discussions}. $\S$\ref{sect:related} presents the related work and $\S$\ref{sect:concl} concludes this paper, respectively.

\section{Background: Amazon Alexa}
\label{sect:prelim-threat-model}

In this section, we introduce Amazon Alexa devices and their common voice service model.

\subsection{Alexa Devices}

There are three kinds of Alexa devices: Amazon Echo, Amazon Tap, and Echo Dot. To support voice commands, they connect to a cloud-based Amazon voice service, \emph{Alexa}.
Amazon Echo is the first generation Alexa device. It always stays in a listening mode, so it does not take any voice commands until a voice word ``Alexa" wakes it up. Every time it wakes up, it serves only one voice command and then returns to the listening mode.
It appears as a 9.25-inch-tall cylinder speaker with a 7-piece microphone array.
Amazon Tap is a smaller (6.2-inch-tall), portable device version with battery supply, but has similar functions.
Echo Dot is the latest generation, which is a 1.6-inch-tall cylinder with one tiny speaker. Both the Echo Dot and the Amazon Echo, which require plug-in power supplies, are usually deployed at a fixed location (e.g., inside a room).
%
%
In this work, we focus on the Echo Dot to examine the Alexa voice service. 

\subsection{Alexa Voice Service Model}
\begin{figure}[t]
\centering
\includegraphics[width=0.95\columnwidth]{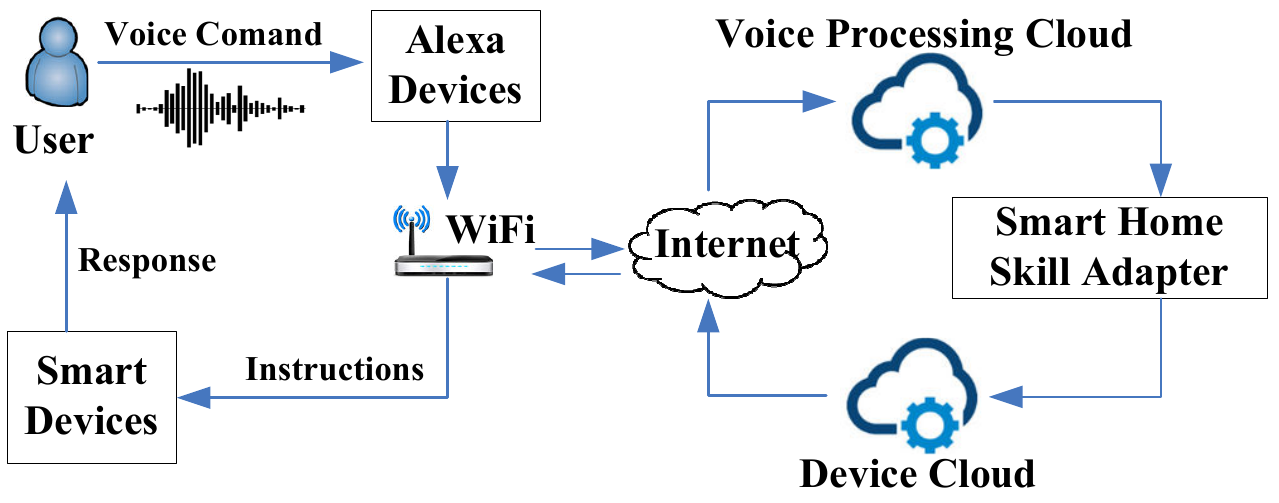}
\caption{Alexa voice service model.}
\label{fig:system}
\end{figure}
The Alexa voice service supports the recognition of voice commands to Alexa devices.
Figure~\ref{fig:system} illustrates how the voice service works with Alexa devices to control smart home devices (e.g., smart bulb, thermostat, etc.).
To control a smart device, a user can speak a voice command to an Alexa device after waking it up with voice ``Alexa".
The Alexa then sends the sounds of that voice command to a remote voice processing cloud via its connected WiFi network.
Once the cloud recognizes the sounds as a valid command, it is forwarded to a server, called \emph{smart home skill adapter},
which is maintained by Amazon to enable the cooperation with third-party service providers.
Afterwards, that command is sent to another cloud which can control the corresponding smart device remotely.
Note that in addition to the control of smart devices, some functions (e.g., checking the weather, placing orders on Amazon.com, etc.) provided by Alexa devices can also be accessed by voice commands.

\section{Breaking Access Control in Alexa}
\label{sect:insecure-voice-auth}

Alexa voice service enables Alexa devices to receive voice commands. However, it has insecure access control so that malicious voice commands may be accepted. More threateningly, they may command smart home devices to assist in crimes. We here identify two vulnerabilities related to the Alexa's access control and one vulnerability existing in the Alexa-enabled smart devices.
We then propose two proof-of-concept attacks. In the following, we start with our threat model.


\begin{table*}[t]
\resizebox{1\textwidth}{!}{
\centering
\scriptsize
\renewcommand{\arraystretch}{1}
\begin{tabular}{|l|l|c|c|}
\hline
      \textbf{Race}        &         \textbf{Age}         & \textbf{Gender}  & \textbf{Result} \\
\hline
\hline
                          & \multirow{2}{*}{10-30}  & F   & \Checkmark \\
\cline{3-4}
      White               &                         & M   & \Checkmark \\
\cline{3-4}
\cline{2-4}
      African American    &  \multirow{2}{*}{30-50} & F    & \Checkmark \\
\cline{3-4}
      Asian               &                         & M   & \Checkmark \\
\cline{2-4}
      Hispano             &  \multirow{2}{*}{50-70} & F   & \Checkmark \\
\cline{3-4}
                          &                         & M    & \Checkmark \\
\hline
\end{tabular}
\centering
\scriptsize
\renewcommand{\arraystretch}{1}
\begin{tabular}{|l|c|c|c|c|c|c|}
\hline
       \textbf{Machines } &  \textbf{Speech Speed} &  \textbf{Alice} & \textbf{Daisy} & \textbf{George} & \textbf{Jenna}  & \textbf{John}  \\
\hline
\hline
       Laptop, Desktop  &  Slow         &  \Checkmark &\Checkmark &\Checkmark &\Checkmark &\Checkmark \\
\cline{2-7}
       Mp3 Player, Bluetooth Speaker  &  Medium         &  \Checkmark &\Checkmark &\Checkmark &\Checkmark &\Checkmark \\

\cline{2-7}
       Home Theater System &  Fast         &  \Checkmark &\Checkmark &\Checkmark &\Checkmark &\Checkmark \\
\cline{2-7}
       Smartphone, Tablet &  Very Fast        &  \Checkmark &\Checkmark &\Checkmark &\Checkmark &\Checkmark \\
\cline{2-7}

\hline
\end{tabular}
}
\caption{Voice command recognition results of Alexa (Left: human sounds; Right: text to speech service)}
\label{tbl-human-synthetic-voice-recognition}
\end{table*}

\smallskip
\noindent\textbf{Threat Model.}
The adversary has no access to the voice processing cloud, the smart home skill adapter, the device cloud, and smart devices.
Victims are the owners of Alexa devices.
The adversary do not require any physical access to the victims' Alexa devices or to be physically present nearby them. They may need to compromise indoor acoustic devices (e.g., Google Chromecast~\cite{HackChromeCast}, answer machines~\cite{HackAnswerMachine}, or Bluetooth speakers~\cite{HackBluetoothSpeaker}) to play arbitrary sounds at victims' home, but do not require to record, overhear, or replay victims' spoken voice commands.
\add{Note that our threat model mainly considers stealthy attacks from the adversary,
We do not consider invasive attack (where an adversary invades the victim's room) and crowd attacks ( multiple adversaries attack at the same time), because too bigger motions or too many adversaries may catch others¡¯ attention.}

We bear in mind that some feasibility tests and attack evaluations might be harmful to Alexa users.
We thus conduct this study in a responsible manner through two measures.
First, for all of our experiments, we use only our own Alexa devices, Alexa service accounts and Alexa-enabled smart devices.
Second, all of the victims are the authors of this work. Note that we seek to disclose new security vulnerabilities of the Alexa devices and services, as well as to devise effective attacks, but not to aggravate the damage.

\subsection{(V1) Weak Single-factor Authentication}
\label{subsect:v1}

Alexa employs a single-factor authentication method based on a password-like voice word, to authenticate users who intend to access the voice service. Nevertheless, the options of that voice word are very limited. By default, it is ``Alexa". In the Alexa mobile application or the Alexa website, users can change it to either ``Amazon", ``Echo", or ``Computer." Anyone who speaks the authentication word of an Alexa device can queue up his/her requests in the device.
Therefore, a malicious user may be able to command the device by obtaining the authentication word through an exhaustive search of those four possible options.
Note that the authentication word is also used to turn an Alexa device out of its listening mode and request it to accept a voice command.

Moreover, Alexa voice service does not support voice authentication (i.e., being able to recognize its owners' sounds), so no matter who speaks voice commands, it accepts all of those which are semantically right and follow correct authentication words.
Given that the Alexa voice commands are publicly known, such weak single-factor authentication allows a fraudulent transaction to be easily made on a victims' Alexa device by its authentication word plus an Alexa command.



\smallskip
\noindent\textbf{Validation.} 
We validate that a valid voice command, which follows a correct authentication word, can be always accepted by an Alexa device, no matter where the voice comes.
We consider voice from both human sounds and text to speech (TTS) service with natural sounding voices. For each of them, we examine several voice variations to show that the Alexa voice service is very powerful to recognize voices. On the other hand, it can be easily abused by the adversary given that weak authentication. Note that the feasibility of the TTS service can make the exploitation of this vulnerability easier, since the voice of a fraudulent command can be generated by its text instead of real human sounds.

For the test of human sounds, we recruit volunteers with different races, ages, and genders. They are requested to speak an authentication word plus a voice command, ``Alexa, What day is today?'' to an Alexa device, Amazon Echo Dot.
It is observed that all the voice commands work for the device; that is, all the volunteers can receive correct answers with respect to their voice commands.
The results are summarized in Table~\ref{tbl-human-synthetic-voice-recognition}.

For the test of the TTS service, we generate a set of Mp3 audio files based on the same voice command with various voice types and speeds from
\texttt{FromTextToSpeech.com}.
We play these audio files to the Alexa device using a variety of machines including computers (e.g., laptop and desktop), music/audio players (e.g., MP3 player, Bluetooth speaker, and home theater system), and mobile devices (e.g., smartphones and tablets). All the voice commands are correctly recognized, and the results are summarized in Table~\ref{tbl-human-synthetic-voice-recognition}.



\smallskip
\noindent\textbf{Causes and lessons.}
The reason that the Alexa device relies on only a weak single factor, but not a stronger one or multiple ones, for its authentication, is two-fold. First, it has limited capabilities of accepting more or arbitrary authentication voice words. Different from voice commands, which are remotely recognized by the voice processing cloud, the authentication voice words are recognized locally in the device. The Alexa device listens for all the surrounding sounds to wait for its authentication voice word in the listening mode. If all the sounds observed over time need to be processed by the cloud, it would result in too much operation overhead. 

Second, it may be too challenging or expensive to support a second factor on voice-related information (e.g., voice authentication, which can recognize each Alexa device's owner voice.). To prevent the Alexa device from being abused by non-owner voices, the second-factor authentication may rely on other kinds of information which can be provided effortlessly by device owners.




\subsection{(V2) No Physical Presence based Access Control}
\label{subsect:v2}

Alexa voice service is designed for the scenario that a user nearby his/her Alexa device speaks voice commands to request services from the device. Since it does not have physical presence based access control, the device can still accept voice commands even if no people are around it. It works for all the sounds reaching it at the sound pressure level (SPL) 60dB or higher. 
Therefore, the sounds from an adversary outside the owner's space or a speaker device may successfully deliver malicious voice commands to the Alexa device.


\smallskip
\noindent\textbf{Validation.}
We validate this vulnerability by testing whether the sounds from a speaker device can successfully queue up requests in an Alexa device where nobody is nearby. The speaker device we use for the test is a Belkin Bluetooth speaker. We
use a smartphone to control it via Bluetooth and play sounds on it from audio files, which we prepare based on several voice commands using the TTS system.
We do the test in various distances between the speaker and the Alexa device: from 2 meters to 12 meters. Note that the speaker's volume is set to the maximum.

Our finding is that the Alexa device accepts our voice commands from the speaker when it is deployed within 8 meters. It is because the speaker's maximum volume can make the sounds arriving at the Alexa device to be at least 60dB only when their distance is not larger than 8 meters.
We further observe that the Alexa can still accept all the voice commands when the sounds are generated by a person instead of the speaker in various distances.

%

\smallskip
\noindent\textbf{Causes and lessons.}
The Alexa device is designed for smart home and mainly deployed in a closed home space, so the assumption of its usage scenario could be that only home members' sounds can reach the device. However, there exist some exceptions. The sounds accepted by the Alexa may come from an adversary outside the home space or a nearby speaker device, which may be compromised and remotely controlled by an adversary.
To prevent such threats, the Alexa device should enforce an access control of accepting voice commands. Physical presence based access control can be a solution, where the device verifies whether any user is present nearby before accepting a voice command.


\subsection{(V3) Insecure Access Control on Alexa-enabled Device Cloud}
\label{subsect:v3}

Alexa owners can control Alexa-enabled smart devices by speaking their names (e.g., ``My Door'') and commands (e.g., ``Open'') via Alexa devices.  Most vendors allow the devices' default names to be replaced, but it is usually not mandatory.  In addition, the commands are always the same for a series of devices from a vendor. It is thus easy for the adversary to guess the voice command of controlling a smart device.
Since the device cloud accepts all the voice commands sent from the Alexa, the security threats caused by Alexa devices may propagate to them. This insecure access control can aggravate the damage caused by the Alexa's vulnerabilities.



\smallskip
\noindent\textbf{Validation.} To validate this vulnerability, we surveyed whether Alexa-enabled smart devices can be controlled via an Alexa device by their default voice names and commands. Table~\ref{tbl-Default-Name} shows a list of feasible smart devices, which are popular, but the vulnerable devices are not limited to them.

%


\begin{table}[t]
\resizebox{0.95\columnwidth}{!}{
\centering
\scriptsize
\renewcommand{\arraystretch}{1}
\begin{tabular}{|l|l|c|c|}
\hline
\textbf{Devices}           &     \textbf{Vendors}          &  \textbf{Default Names}    \\
\hline
\hline
Garage Door       &     Garageio        &  My Door          \\
\hline
Smart Bulb        &      Tp-Link        &  My Smart Bulb   \\
\hline
Smart Plug        &     TP-Link         &  My Smart Plug    \\
\hline
Smart Switch      &     WeMo            &  WeMo Light Switch   \\
\hline
Learning Thermostat  &    Nest          &  Thermostat \\
\hline
\end{tabular}
}
\caption{Alexa-enabled smart devices can be controlled by their default voice names and commands.}
\label{tbl-Default-Name}
\end{table}

\smallskip
\noindent\textbf{Causes and lessons.}
This vulnerability lies in that the device cloud relies on the Alexa for its access control. It assumes that the Alexa has a secure authentication method and all the voice commands sent from the Alexa are benign. However, the Alexa's current authentication may be breached.
To reduce the risks of being abused, the Alexa-enabled device cloud may need to have a light-weight security mechanism (e.g., a secret short code and mandatory default name change).

\subsection{Proof-of-concept Attacks}
\label{subsect:proof-of-attack}

We now devise two practical attack cases based on the above vulnerabilities: home burglary and fake order.
Both of them can cause Alexa owners to suffer from a large financial loss.
To launch these attacks, the adversary needs to spread his/her malware to discover suitable victims whose Alexa devices can be abused to receive fraudulent voice commands without the adversary's nearby presence.
Note that our attack cases are used for the victims who are identified to be vulnerable by the malware, but not to target specific victims. Even though the percentage of possible victims to all Alexa owners is low, not only can those victims get unexpected damage, but also are the other owners still under the risk of the crimes.
Before presenting the attacks, we seek to answer two major questions.

%
\begin{itemize}
\item How can an Alexa device be abused without being present nearby?
\item How to enable victims discovery in the malware?
\end{itemize}

In the following, we present some example cases to show the attack feasibility, but the available avenues are not limited to them.

\subsubsection{Abusing an Alexa device without being Present Nearby}

One possible way to deliver sounds to an Alexa device without being present nearby is to exploit an indoor acoustic device which is placed in the same room. It can also be achieved by broadcasting from the radio or a fire fighting speaker in a house or a building. We here focus on the acoustic device.

Nowadays, there have been many indoor acoustic devices, e.g., Roku, Chromecast, Chromecast Audio, smart TVs, Bluetooth speakers, Apple TV box, etc. Some of them can be compromised by the adversary to deliver sounds to Alexa devices, once they are deployed in the same indoor space. To illustrate the abuse, we take both Bluetooth speakers and smart TVs as examples.

\smallskip
\noindent\textbf{Bluetooth speakers.}
It is not rare that a Bluetooth speaker~\cite{Buletooth} and an Alexa device coexist in the same space, due to two reasons.
First, it is recommended to connect an Alexa device to a Bluetooth speaker so as to enjoy high-quality audio. Second, Bluetooth speakers are increasingly popular with ongoing ``wirelessization" in smart homes.
The adversary outside the victim's house can connect his/her smartphone to the victim's Bluetooth speaker and then plays an MP3 audio file of voice commands. To show the viability, we use a Nexus smartphone and a Belkin Bluetooth speaker, as well as generate the audio using the TTS system, in our experiments. It is observed that the Alexa takes the voice commands from the audio and acts accordingly.


\smallskip
\noindent\textbf{Smart TVs.}
Smart TVs, which are getting popular in recent years, support to play streaming videos from Youtube.com. A user is allowed to cast a specific Youtube video from his/her mobile device (e.g., smartphone, tablet, etc.) to a smart TV through the home WiFi.
The adversary can steal the victim's home WiFi password through the malware without root permission~\cite{RecoverWiFiRouter,DefaultPassofWirelessRouter}, and then cast a video with voice commands to the victim's smart TV when next to his/her house. We confirm the viability using an LG smart TV.

\begin{figure}[t]
\centering
\includegraphics[width=0.85\columnwidth]{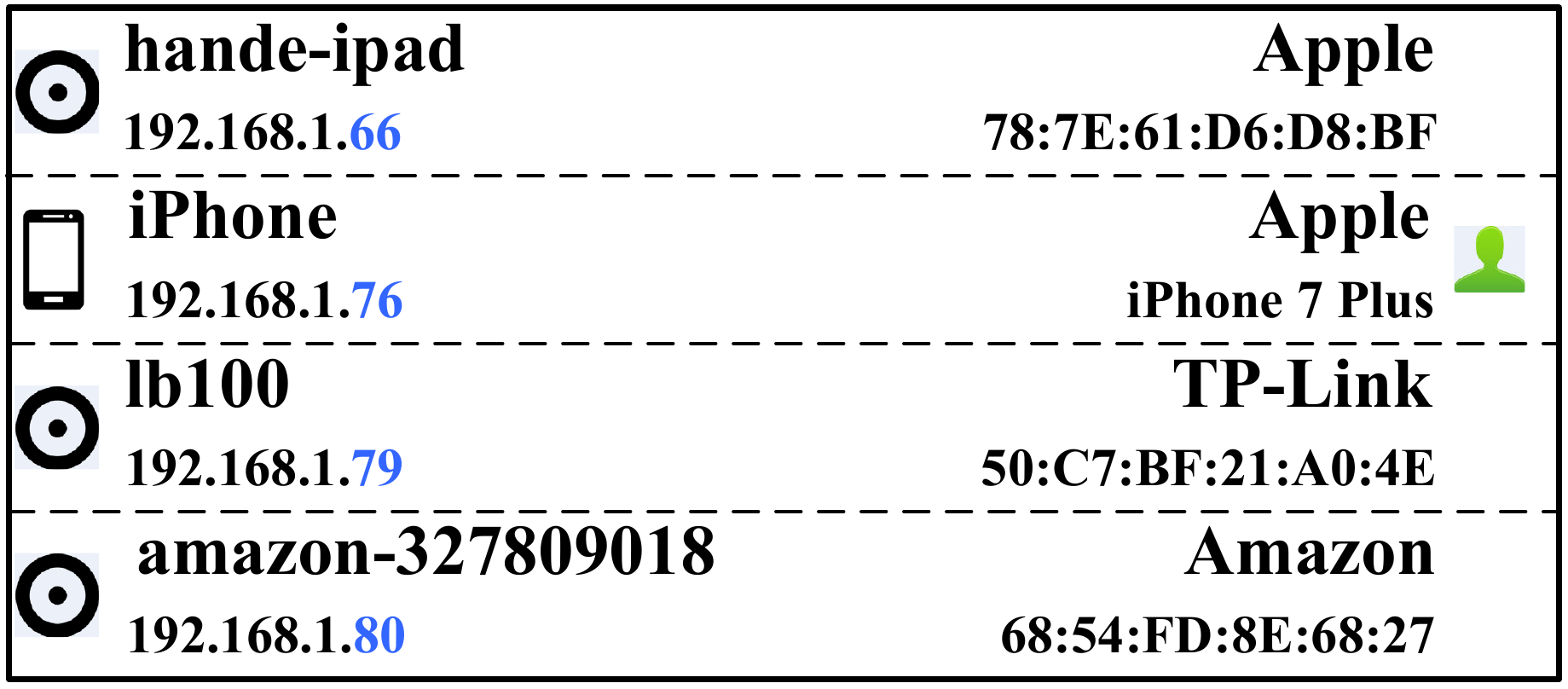}
\caption{The result of a WiFi scan.} \label{fig:Scan}
\end{figure}

\subsubsection{Discovering victims using the Malware}
The adversary can discover potential victims, which have deployed Alexa devices or/and Alexa-enabled smart devices at home, using the smartphone malware without root permission.
Since both the Alexa devices and the smart devices need to connect to a WiFi network, the malware can scan the home WiFi network with which its smartphone host associates to detect whether they exist or not. We find that their device names have the same patterns. Specifically, the Alexa devices are named in the pattern of ``amazon-X'', where ``X'' represents product IDs.
We take a smart bulb as an example for the Alexa-enabled device. It has a name ``lb100'', which stands for a 100-watt light bulb. We validate this WiFi scan approach using a free Android application~\emph{Fing}, which scans WiFi devices with only the permissions of Internet access and reverse DNS.
The snapshot of the WiFi scan result by using Fing is shown in Figure \ref{fig:Scan}.

Note that in order to launch attacks, the adversary requires to know where the victim's home is. The malware can report it by querying location information from mobile OS APIs, e.g., Android localization interface~\texttt{LocationListener}.

\subsubsection{Two Attack Cases}

We now describe two attacks based on the Alexa's vulnerabilities.

\smallskip
\noindent\textbf{Home Burglary.} An adversary can burglar the house of an Alexa device owner by requesting the Alexa to open a door via an Alexa-enabled smart lock.
Note that there are many Alexa-enabled smart devices, which include various locks, lamps, etc.
Most of them support the default voice commands of the Alexa service, so they can be easily abused once a fraudulent request can be made on the Alexa device. For example, an adversary can open a victim's Garageio garage door by using the default voice command, ``Alexa, tell Garageio to open my door''~\cite{Garageio}.


\smallskip
\noindent\textbf{Fake Order.}  The Alexa service may be abused to place a fake order of a device owner by the adversary, and then the owner may suffer from financial loss. For example, shopping on Amazon.com is one of popular Alexa services. Once the adversary can deliver sounds to the Alexa device, s(he) can place an order of any prime-eligible items (e.g., Dell Inspiron 5000 laptop at \$899.99), which are charged on the owner's Amazon account, using the voice command, ``order [item name].'' Since the ordered items are by default delivered to the owner's home address within two days because of the Amazon Prime shipping service, the adversary can stealthily pick them up in front of the victim's house. Note that we have verified that for this kind of orders, Amazon carriers usually leave the ordered items to the front door without the need of a signature.

\section{Virtual Security Button (\texttt{VSButton}): Physical Presence based Access Control}
\label{sect:sol}

We propose an access control mechanism which is based on physical presence to Alexa devices or other devices/services that require the detection of physical presence. It not only addresses V2, but also makes Alexa's authentication to be two-factor instead of current single factor (i.e., V1).
We name this mechanism as virtual security button (\texttt{VSButton}), because whether physical presence is detected is like whether a virtual button is pushed. The access to a device/service with \texttt{VSButton} is not allowed when the virtual button is not in a push state (i.e., physical presence is not detected).
As a result, this mechanism enables Alexa devices to prevent fraudulent voice commands which are delivered when no persons are nearby them.

The mechanism detects physical presence based on the WiFi technology. It results in negligible overhead on the Alexa device/service, because it reuses the user's existing home WiFi network and needs negligible change on how the user requests Alexa services. It does the detection by monitoring the channel state information (CSI) of the channel used by the home WiFi network. The CSI changes represent that some human motions happen nearby the Alexa device.
Once any human motion is detected, the Alexa device is activated to accept voice commands.
Therefore, the user just needs to make a motion (e.g., waving a hand for 0.2 meters) to activate the device before speaking his/her voice commands.




We believe that detecting human motions based on WiFi signals is a practical yet low-cost solution approach due to two reasons.
First, home WiFi networks are commonly deployed, so no extra deployment cost is needed.
Second, only a software upgrade is required for the Alexa devices, since all of them have been equipped with WiFi.
Before presenting our \texttt{VSButton} design, we introduce the CSI primer and the CSI-based human motion detection, which is based on the multi-path/multi-reflection effects.

\subsection{CSI Primer}
\label{subsect:primer-of-csi}

It is commonly used to characterize the channel state properties of WiFi signals.
Current WiFi standards like IEEE 802.11n/ac rely on the Orthogonal Frequency Division Multiplexing (OFDM) technique, which divides a channel into multiple subcarriers, and use the Multiple-Input Multiple-Output (MIMO) technology to boost speed.
Each CSI value represents a subcarrier's channel quality (i.e., channel frequency response) for each input-output channel.

%
%

The mathematical definition of the CSI value is presented below.
Let $\mathbf{x}_i$ be the $N_T$ dimensional transmitted signal and $\mathbf{y}_i$ be the $N_R$ dimensional received signal in subcarrier number $i$. 
For each subcarrier $i$, the CSI information $\mathbf{H}_i$ can be obtained based on the following equation:
\begin{equation}\label{eq10}
\mathbf{y}_i=\mathbf{H}_i\mathbf{x}_i+\mathbf{n}_i,
\end{equation}
where $\mathbf{n}_i$ represents noise vector.
Without considering the noise, it can be known from Equation (\ref{eq10}) that the matrix $\mathbf{H}_i$ transfers the input signal vector $\mathbf{x}_i$ to be output signal vector $\mathbf{y}_i$.
It gives us the intuition about why the channel information is encoded in (and hence represented by) the matrix $\mathbf{H}_i$.
Commodity WiFi devices can record thousands of CSI values per second from numerous OFDM subcarriers, so even for subtle CSI variations caused by a motion, the CSI values can still provide descriptive information to us.
%

\begin{figure}[t]
\centering
\includegraphics[width=0.9\columnwidth]{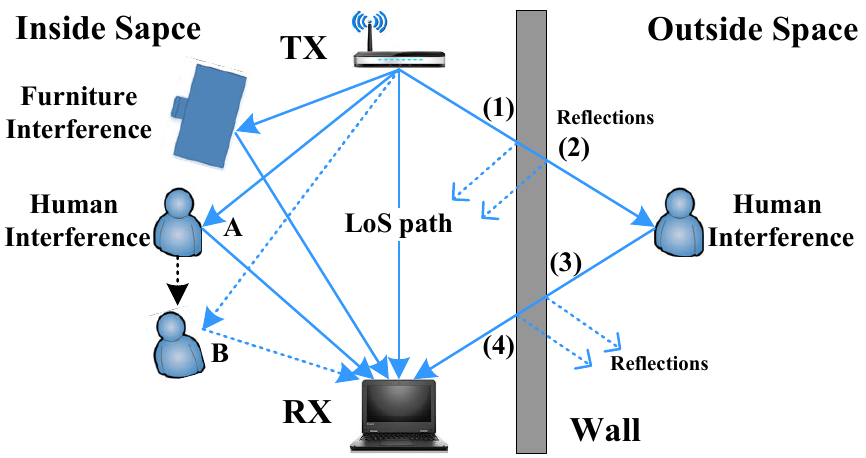}
\caption{An illustration of multi-path and multi-reflection effects.} \label{Multipath}
\end{figure}

\subsection{CSI-based Human Motion Detection}

We detect whether there are any human motions and identify whether they happen inside a house/room by leveraging the multi-path and multi-reflection effects on CSI, respectively.

\smallskip
\noindent
\textbf{Multi-path Effect for Human Motions Detection.}
The multi-path effect refers to the signal propagation phenomenon that a wireless signal reaches a receiving antenna along two or more paths. As shown in Figure~\ref{Multipath}, the receiver (RX) receives multiple copies from a common signal along multiple paths,
the line-of-sight (LoS) path, one reflection from the human at location A, and one reflection from the furniture. Different lengths of the paths along which the WiFi signals are sent result in phase changes of the signals, thereby leading to various CSI values.
Therefore, human motions can change the paths of WiFi signals and further make CSI values to change.
For example, when the human moves from location A to location B, the new signal reflection path is being substituted for the original one. This move can thus distort the CSI values observed at the receiver.

\smallskip
\noindent\textbf{Multi-reflection Effect for Identifying Where the Motions are.} The multi-reflection effect means that a wireless signal may be reflected by multiple objects and it can cause the receiver to receive multiple copies of signals from different reflections. When we consider human motions inside and outside a room/house, they can be differentiated in terms of variation degrees of CSI values due to multiple different reflections.
As shown in Figure~\ref{Multipath}, when a WiFi signal is reflected by the human body outside the wall, the signal arriving at the receiver along this path should have experienced 
four reflections: (1), (2), (3), and (4). They happen since the signal traverses different media, from air to wall and from wall to air. 
Such multi-reflection effect causes the signal received by the receiver to suffer from a serious attenuation.
Our results show that it makes outside motions to result in only a small variation of CSI values, compared with a significant variation caused by inside human motions.
The variation degrees of CSI values can thus be leveraged to identify the human motions occurring inside and outside the wall.

\subsection{\texttt{VSButton} Design}
\label{subsect:vsbutton-design}
\begin{figure}[t]
\centering
\includegraphics[width=0.9\columnwidth]{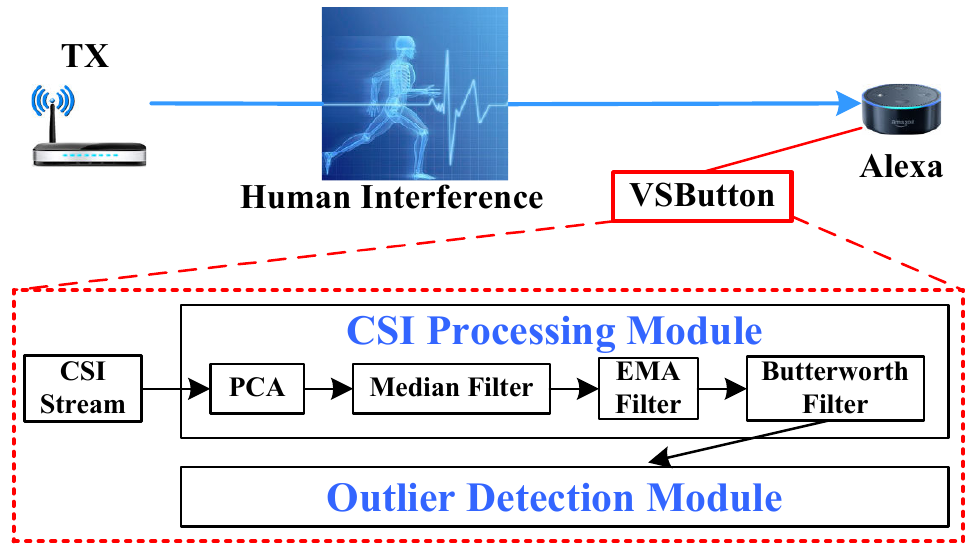}
\caption{VSButton design.} \label{vsbutton}
\end{figure}

In this section, we introduce our \texttt{VSButton} design as shown in Figure~\ref{vsbutton}. It resides at the Alexa device and monitors human motions by its collected CSI values of the data packets received from the WiFi AP. Based on the CSI variations of the wireless channel between the device and the AP, \texttt{VSButton} can determine whether any human motion happens inside or not. Note that, in order to keep collecting CSI values over time, the Alexa device can send ICMP packets to the AP at a constant rate (e.g., 50 ICMP messages/second in our experiments) and then keep receiving the packets of ICMP reply messages from the AP.

The detection of inside human motions in the \texttt{VSButton} mainly consists of two phases, \emph{CSI Processing Phase} and \emph{Outlier Detection Phase}. When receiving CSI values, the former eliminates noises from them so that the CSI variation patterns caused by human motions can be augmented.
Based on the output of the first phase, the latter relies on a real-time outlier detection method to detect the CSI patterns of the human motions inside a room/house.
In the following, we present the details of each phase and then give an example of identifying indoor human motions.

\subsubsection{CSI Processing Phase}
%
This phase consists of three modules: principal component analysis (PCA)~\cite{jolliffe2002principal}, median and exponential moving average (EMA) filter~\cite{arce2005nonlinear, holt2004forecasting}, and Butterworth low-pass filter~\cite{butterworth1930theory}. We first use the PCA module to reduce the dimensions of CSI values by removing those uncorrelated to motions detection. The last two modules are used to eliminate bursty noise and spikes, and filter out high-frequency noise in CSI stream values, respectively.

\begin{figure}[t]
\centering
\includegraphics[width=0.9\columnwidth]{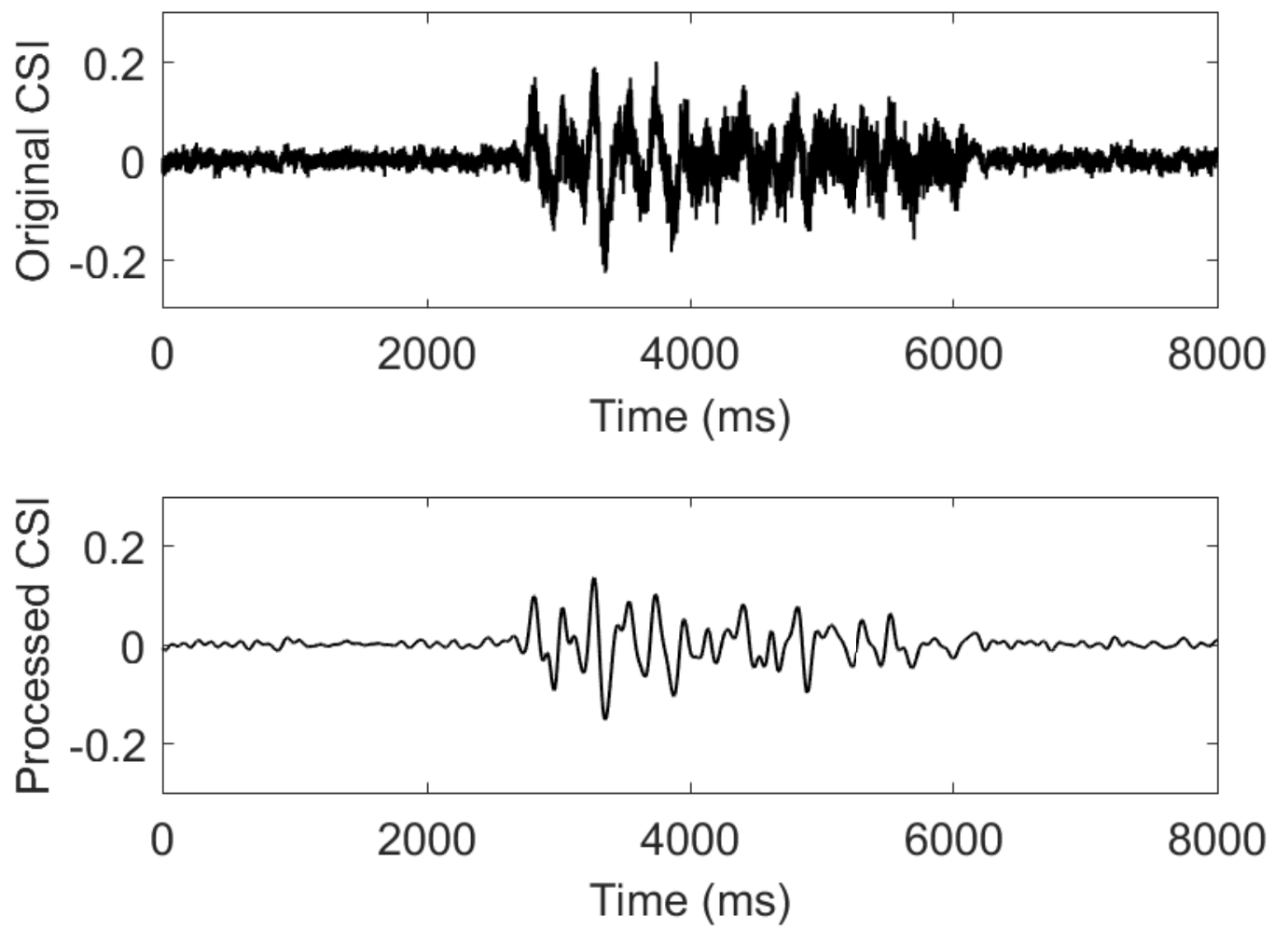}
\caption{Comparison between original/processed CSI over time.} \label{comparison}
\end{figure}

Figure \ref{comparison} gives an example of the comparison between the original CSI over time and the CSI which has been processed by those three modules.
It is shown that most of the noises in the original CSI are removed so that the processed one can give us more accurate information for motions detection.
We elaborate the details of each module below.

\smallskip
\noindent\textbf{PCA module.}
We employ the PCA to remove uncorrelated noisy information from the collected CSI by recognizing the subcarriers which have strong correlations with motions detection. The PCA is usually used to choose the most representative principal components from all CSI time series. It can accelerate the subsequent signal processing because the collected CSI may contain too much noisy information.
Take an Intel Link 5300 WiFi adapter, which has $N_T$ transmit antennas and $N_R$ receive
antennas, as an example. Each transmit-receive channel has 30 subcarriers, so there are $N_T\times N_R\times 30$ CSI streams to be generated.
%
%
%

Since the collected CSI information can be represented by a high dimensional matrix, the PCA can be done by the following five components: 1) data centralization, 2) calculation of covariance matrix, 3) calculation of eigenvalues and eigenvectors, 4) selection of main eigenvalues, and 5) data reconstruction. The PCA details can be found in \cite{abdi2010principal}.
In our experiments, it is observed that the first four components almost show the most significant changes in CSI streams but the first one is too sensitive to signal noise. As a result, we keep only the second, the third, and the forth components for further analysis.

%
%

\smallskip
\noindent\textbf{Median and EMA Filter Module.}
We next use a combination of a median filter and an EMA filter to eliminate bursty noise and spikes in CSI stream values.
They happen because commodity WiFi interface cards may have a slightly unstable transmission power level and also be affected by dynamic channel conditions (e.g., air flow, humidity, etc.).
The median filter is often used to remove noise from the signal.
Its main idea is to run through the signal entry by entry while replacing each entry with the median of neighboring entries.
It can smooth out short-term fluctuations and highlight long-term trends.
Note that the number of neighboring entries (i.e., moving window size) is a configurable parameter.
We also adopt the EMA filter to smooth CSI values. It applies weighting factors which decrease exponentially to each older CSI value.  The window size of EMA is a configurable parameter.

\smallskip
\noindent\textbf{Butterworth Filter Module.}
We finally apply the Butterworth low-pass filter to filtering out high frequency CSI, since human motions cannot be generated too fast. Specifically, it is observed that the CSI variations caused by human motions mainly happen in the low frequency domain (i.e., less than 100~Hz). 
Given a proper cut-off frequency, 100~Hz, high frequency noise can be removed by the filter.

%

\subsubsection{Outlier Detection Phase}
We detect human motions using a real-time hyper-ellipsoidal outlier detection method over non-stationary CSI data streams~\cite{moshtaghi2011incremental}. It improves accuracy over the typical moving average method, which detects an anomaly based on a threshold of the distance between current value and the average of old values, from two aspects. First, it employs a new distance metric, i.e., \emph{Mahalanobis distance}, which is more accurate. Second, it exploits exponential moving average (EMA) to update the previous sample mean.

%
%

This detection method can be mathematically described as follows. Let $X_k=\{x_1, \cdots ,x_k\}$ be the first $k$ samples of CSI readings.
Each sample is a $d\times 1$ vector, where $d$ is the number of chosen components.
The sample mean $m_k$ and sample covariance $S_k$ are given by
\begin{equation}
m_k=\frac{1}{k}\sum_{i=1}^{k}{x_i},~~ S_k=\frac{1}{k-1}\sum_{i=1}^{k}(x_i-m_k)(x_i-m_k)^{T}.
\end{equation}
The Mahalanobis distance of a sample reading $r$ from the $X_k$ is defined as
\begin{equation}\label{eq-20}
D(r, X_k)=\sqrt{(r-m_k)^{T}S_{k}^{-1}(r-m_k)}.
\end{equation}
By using Mahalanobis distance, we consider the reading $r$ as an anomaly if $D(r, X_k)>t$, where $t$ is a threshold parameter and needs to be carefully selected according to the experiments.
All of the points bounded by $D(r, X_k)\leq t$ are considered as normal readings.
%
%

When we apply it to the non-stationary CSI streams, the sample mean is updated by
\begin{equation}
m_{k+1}=\alpha m_{k}+ (1-\alpha)x_{k+1},
\label{eq-forgetting-factor}
\end{equation}
where $\alpha\in(0,1)$ denotes a forgetting factor.
%
The closer the receiving of a CSI value reading is, the larger weight it has to determine the next sample mean.

For the motions detection, we avoid false detections by specifying a threshold for the number of consecutive anomalous CSI values (10 is used in our experiments).
It is because the noises may occasionally lead to some anomaly detections. However, human motions can make anomalous CSI value readings to last for a long period of time (e.g., consecutive 10 readings).


%
%

\begin{figure}[t]
\centering
\subfigure[Indoor motion CSI variation]{
\label{fig:indoor}
\includegraphics[width=0.8\columnwidth]{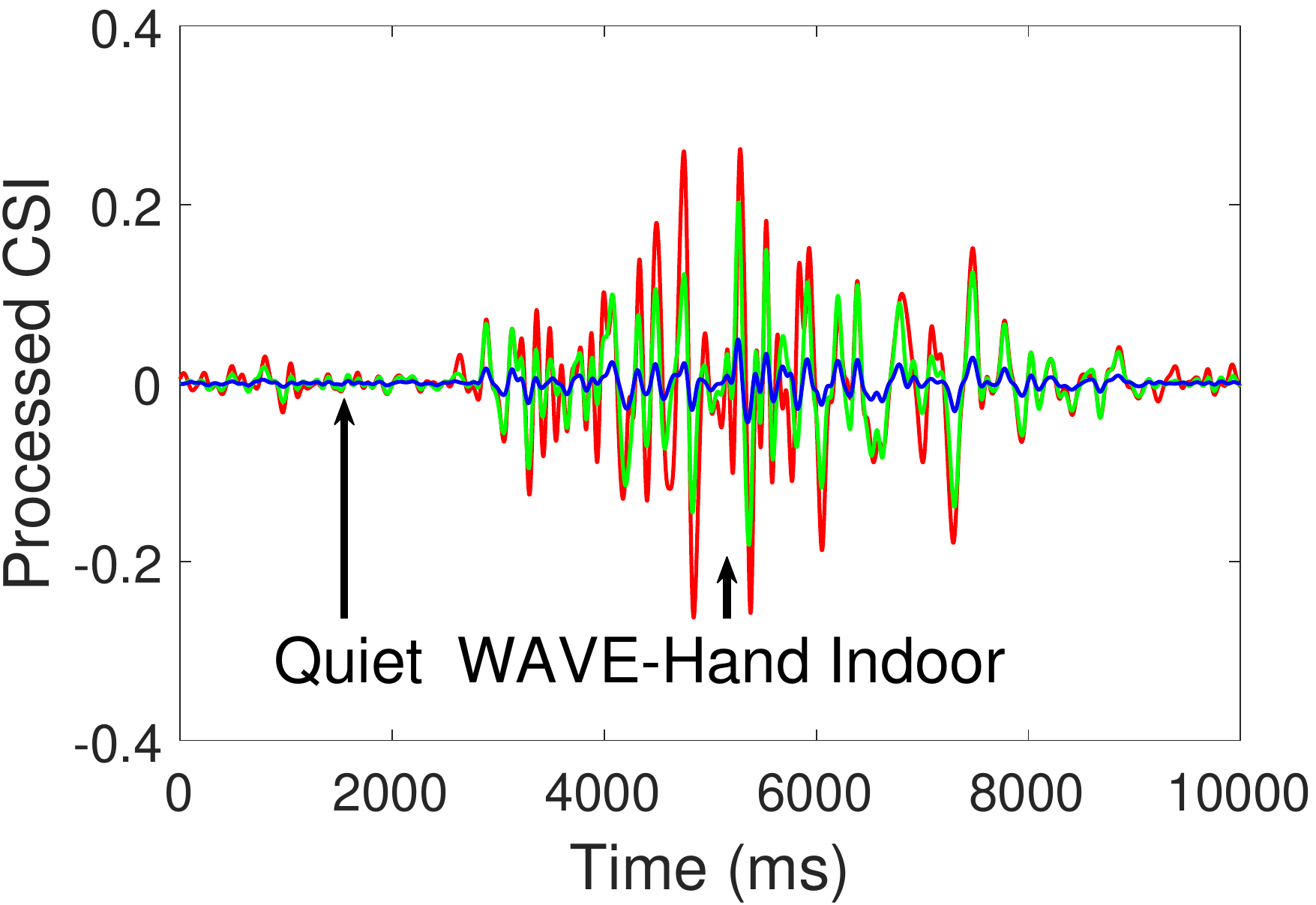}
}
\subfigure[Outdoor motion CSI variation]{
\label{fig:outdoor}
\includegraphics[width=0.8\columnwidth]{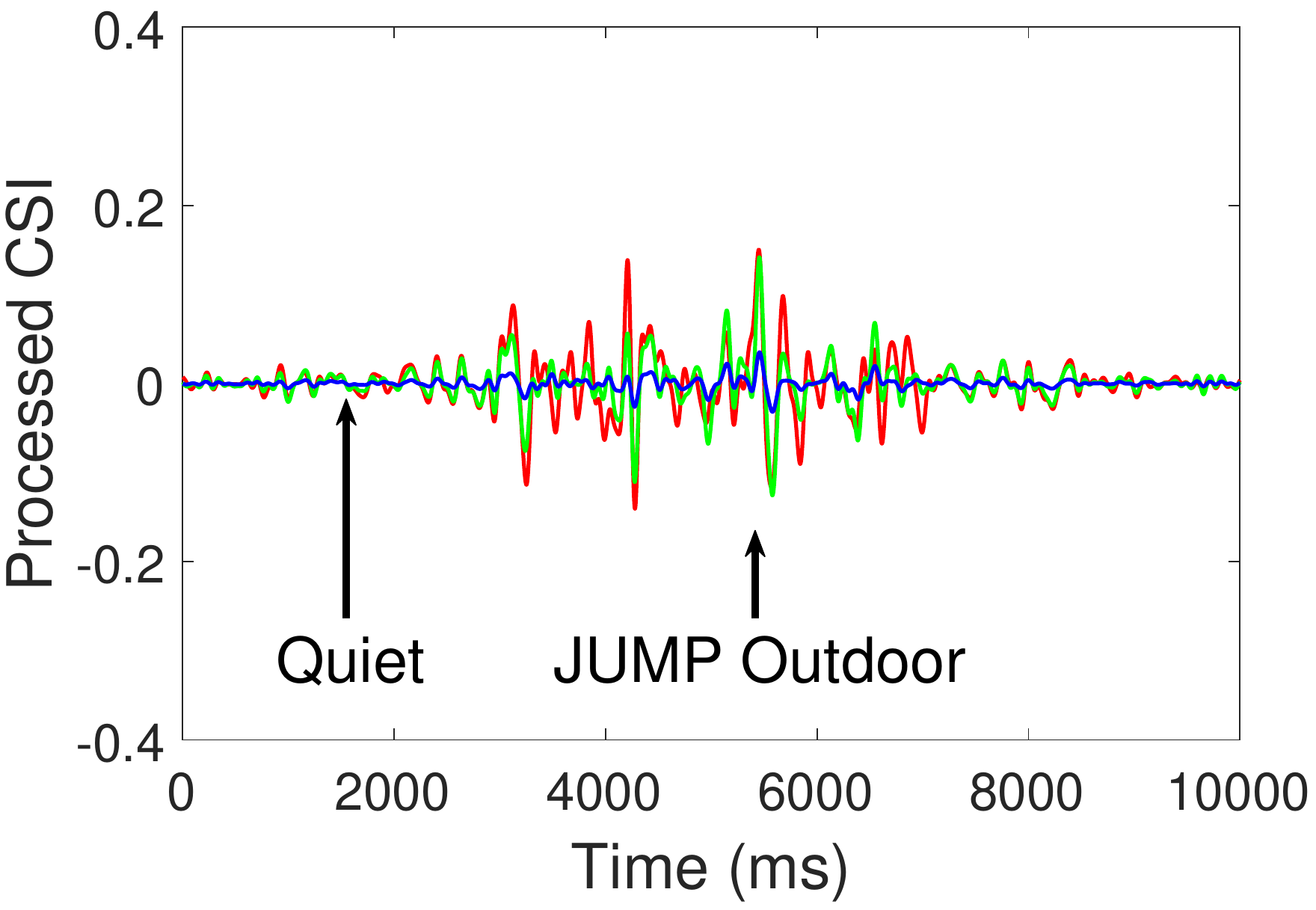}
}
\caption{Comparison of indoor and outdoor CSI variations.}
\label{fig:indoor-outdoor-com}
\end{figure}

%
%

\subsubsection{An Example of Identifying Indoor Motions}
We show that the CSI values of indoor and outdoor motions can be clearly differentiated so that the indoor motions can be properly detected. Figures~\ref{fig:indoor} and ~\ref{fig:outdoor} plot the processed CSI values respectively for a small indoor motion (i.e., waving a hand) in one laboratory room and a large outdoor motion (i.e., jumping). It is observed that the peak CSI values are 0.27 and 0.14, respectively. It shows that even the small indoor motion can lead to the CSI variations which are much larger than those of the strong outdoor motion. We believe that with proper parameter configurations, \texttt{VSButton} is able to correctly identify indoor motions and then activate an Alexa device to accept voice commands.
More experimental results will be given in Section~\ref{sect:eval}.

\section{Implementation and Evaluation}
\label{sect:eval}
In this section, we present the implementation of our \texttt{VSButton} prototype and the evaluation results of both laboratory and real-world settings.


\begin{figure}[t]
\centering
\includegraphics[width=0.8\columnwidth]{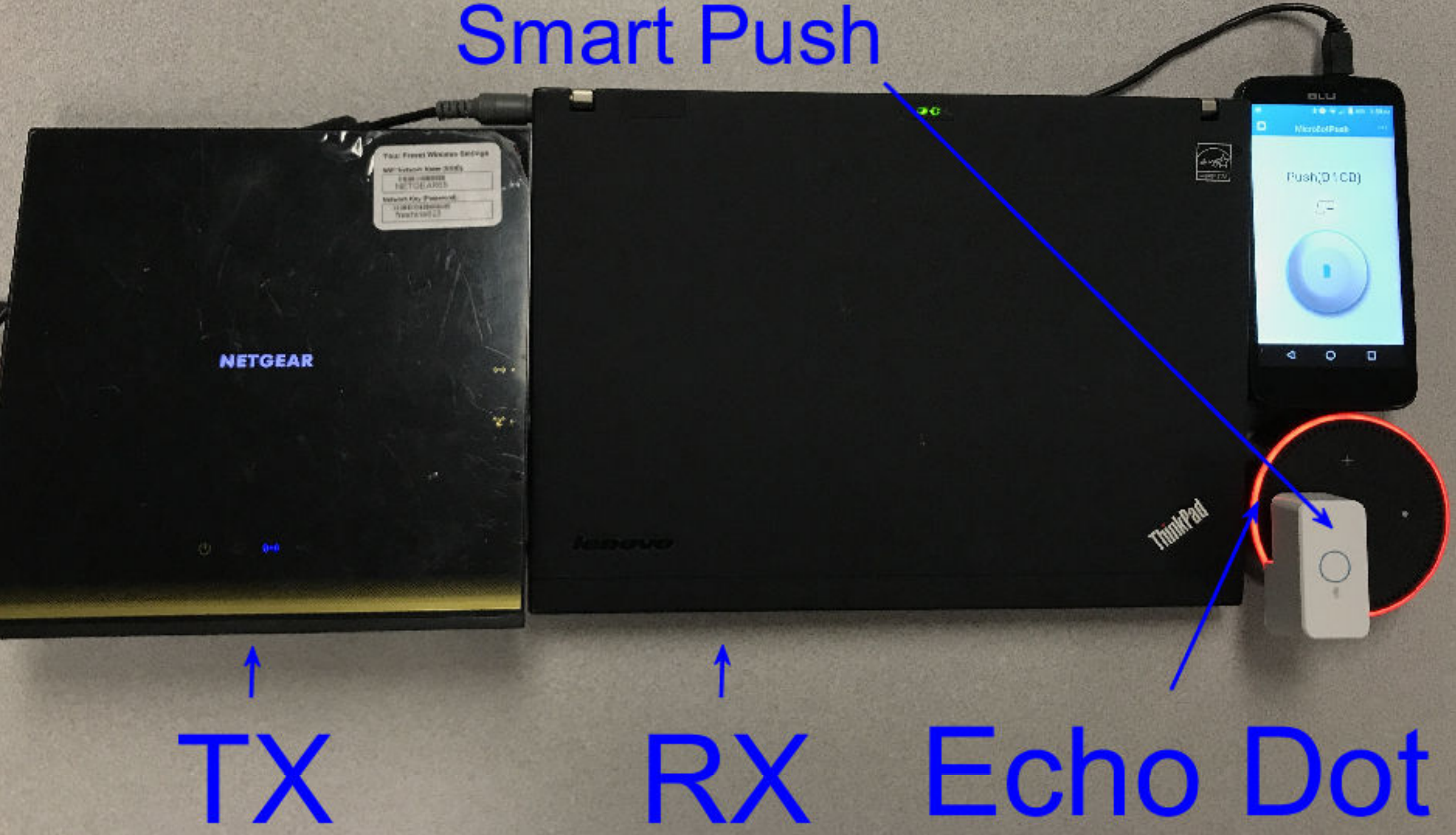}
\caption{\texttt{VSButton} prototype.}
\label{Prototype}
\end{figure}

\subsection{Prototype Implementation}

Our \texttt{VSButton} prototype is based on commercial off-the-shelf (COTS) devices, as shown in Figure~\ref{Prototype}. The TX device is a Netgear R63000v2 WiFi router, which can be considered as a home WiFi AP. It is employed as the transmitter for the packets used by the Alexa device to collect CSI over time. The AP is set to the 802.11n mode~\cite{IEEE802}, because the Alexa devices have not supported 802.11ac yet~\cite{AlexaWifiSpec}.
Since the Alexa devices are not open to development, we implement the motions detection module on a laptop, which is tagged to be RX. The Alexa device, Echo Dot, can then get motions detection result from the laptop.
The RX device is a Lenovo X200 laptop equipped with an Intel Link 5300 WiFi adapter, which is able to collect CSI values using the tool developed by the work~\cite{halperin2011tool}.
To emulate the Alexa device's access control, the module controls MicroBot Push~\cite{MicroBotPush} to turn on/off the Alexa device's microphone through a smartphone, once any status change of access control is detected by the module based on motions detection.

Note that in our current \texttt{VSButton} prototype, there are a wireless router, a laptop, a smartphone, and a MicroBot. Seemingly, the deployment cost is not small. However, most of people have deployed a wireless router at home for Internet access and the others' functions can be integrated to the Alexa devices based on only software upgrades.
\begin{table}[t]
\resizebox{0.95\columnwidth}{!}{
\scriptsize
\centering
\renewcommand{\arraystretch}{1}
\begin{tabular}{|l|l|}
\hline
\textbf{Parameters} &  \textbf{Values} \\
\hline
Sampling rate of CSI values     &   50\\
\hline
Median filter window size     &   9  \\
\hline
EMA filter windows size       &   15 \\
\hline
Cut-off frequency of Butterworth filtering (rad/s)  & $4 \pi~ $\\
\hline
Forgetting factor $\alpha$        &   0.98  \\
\hline
\end{tabular}
}
\caption{Experiment settings}
\label{tble-exp-settings}
\end{table}

\subsection{Evaluation}
\label{subsect:sol-eval}

We next introduce our experimental settings and evaluate the performance of our \texttt{VSButton} prototype in three space settings: \emph{square room}, \emph{rectangle room}, and \emph{two-bedroom apartment}. The performance refers to whether three cases, no motion, indoor motion, and outdoor motion, can be correctly identified. We recruit six volunteers to participate in the experiments. They are required to do three motions including waving a hand (WAVE-HAND), sitting down and standing up (SIT-DOWN-STAND-UP), and jumping (JUMP, 0.5m), inside and outside a room. They represent three degrees of human motions, weak, medium, and strong, respectively. Note that we examine the Mahalanobis distance for each measurement and see whether indoor motions can be clearly detected or not.

\subsubsection{Experimental Settings}
\label{subsect:sol-testbed}

In all experiments, the RX (the laptop with an Echo dot) sends 50 ICMP \texttt{Echo Request} messages per second to the TX (the WiFi router) so that it can keep collecting CSI over time from the ICMP \texttt{Echo Reply} messages sent by the TX.
A CSI stream with the sampling rate of 50 values per second can thus be used for motions detection.
Each ICMP message size is 84 bytes. The collection of the CSI stream requires only around 8.2~Kbps bandwidth.

The windows sizes of the median and EMA filters are set to 9 and 15, respectively. 
Our experimental results show that these two numbers are large enough for the filters to remove noise.
The cut-off frequency~\cite{butterworth1930theory} for the Butterworth filter is set to $\omega_c=\frac{2\pi\times 100}{50} = 4 \pi~ $rad/s\footnote{rad/s is the unit of rotational speed (angular velocity)}, because human motions lead to only low-frequency CSI variations, which are typically less than $f=100$ Hz.
In the outlier detection module, we set the forgetting factor $\alpha$ to be 0.98. It means that we give a larger weight to the recent CSI value readings. 
%
%
The experimental settings are summarized in Table \ref{tble-exp-settings}.

\begin{table*}[t]
\centering
\resizebox{0.95\textwidth}{!}{
\normalsize
\renewcommand{\arraystretch}{1}
\begin{tabular}{|l|c|c|c|c|c|c|c|c|c|c|}
\hline
 \textbf{Square Room}      &  \multicolumn{4}{|c|}{\textbf{Indoor Locations}} & \multicolumn{6}{|c|}{\textbf{Outdoor Locations}}\\
\hline
   locations & $A$  & $B$ & $C$  & $D$ & $A'$ & $B'$ &  $C'$ & $D'$ & $M'$ &  $N'$  \\
\hline
\hline
WAVE-HAND  & $0.218$   & $0.213$ & $0.195$ &\underline{$\mathbf{0.191}$} &  $0.104$  & $0.101$  &  $0.079$ & $0.083$ & $0.156$ & $0.121$    \\
\hline
SIT-DOWN-STAND-UP   & $0.277$ & $0.271$   & $0.258$ & $0.253$ & $0.118$ & $0.113$ & $0.088$ & $0.092$ & \underline{$\mathbf{0.238}$}  & $0.139$  \\
\hline
JUMP  & $0.392$  &  $0.391$ & $0.371$ & $0.366$ &  $0.132$ & $0.128$ & $0.099$ & $0.103$ & \underline{$\mathbf{0.373}$} & $0.165$ \\
\hline
DO NOTHING  & $0.026$  &  $0.021$ & $0.027$ & $0.024$ &  $0.023$ & $0.027$ & $0.028$ & $0.023$ & $0.020$ & $0.023$ \\
\hline
\end{tabular}
}
\caption{Mahalanobis distance measured in a square room with Configuration 1.}
\label{tbl-max-distance-for-squre-room-config1}
\end{table*}

\begin{table*}[t]
\centering
\resizebox{0.95\textwidth}{!}{
\normalsize
\renewcommand{\arraystretch}{1}
\begin{tabular}{|l|c|c|c|c|c|c|c|c|c|c|}
\hline
 \textbf{Square Room}      &  \multicolumn{4}{|c|}{\textbf{Indoor locations}} & \multicolumn{6}{|c|}{\textbf{Outdoor locations}}\\
\hline
   locations & $A$  & $B$ & $C$  & $D$ & $A'$ & $B'$ &  $C'$ & $D'$ & $M'$ &  $N'$  \\
\hline
\hline
WAVE-HAND  & \underline{$\mathbf{0.312}$}   & $0.315$ & $0.401$ & $0.409$ & $0.041$ & $0.043$ & $0.049$  & $0.051$ & $0.092$ & $0.063$\\
\hline
SIT-DOWN-STAND-UP   & $0.345$ & $0.349$   & $0.423$ & $0.430$ & $0.060$ & $0.062$ & $0.069$ & $0.071$ & $0.121$ & $0.089$  \\
\hline
JUMP  & $0.401$  & $0.407$ & $0.451$ & $0.459$ & $0.069$ & $0.071$ & $0.084$ & $ 0.086$ & \underline{$\mathbf{0.241}$} & $0.099$ \\
\hline
DO NOTHING  & $0.025$  &  $0.021$ & $0.022$ & $0.024$ &  $0.028$ & $0.026$ & $0.021$ & $0.022$ & $0.023$ & $0.025$ \\
\hline
\end{tabular}
}
\caption{Mahalanobis Distance measured in a square room with Configuration 2.}
\label{tbl-max-distance-for-squre-room-config2}
\end{table*}
\normalsize

\subsubsection{A Square Lab Room}
\begin{figure}[t]
\centering
\subfigure[Configuration 1]{
\label{SR1}
\includegraphics[height=3cm]{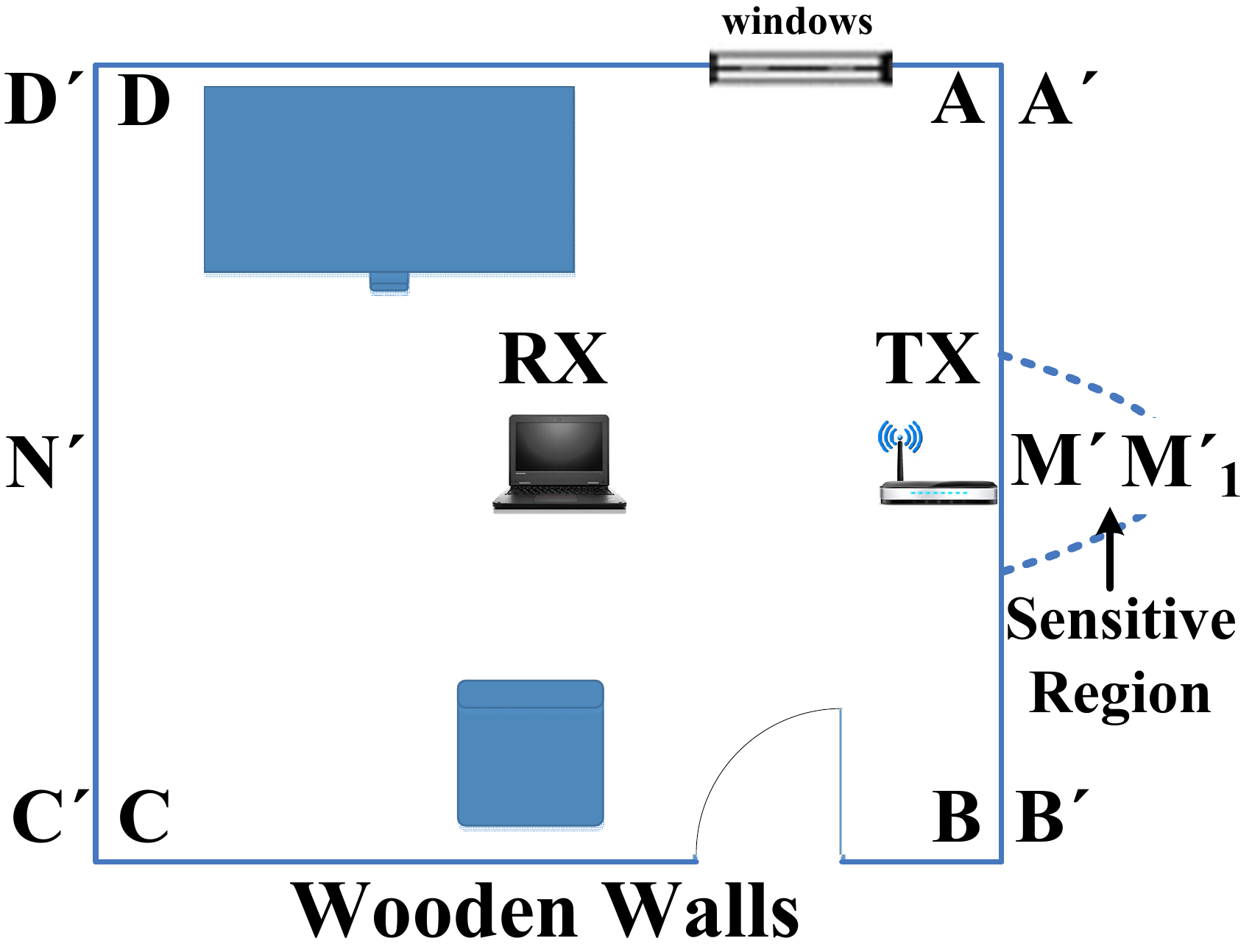}
}
\subfigure[Configuration 2]{
\label{SR2}
\includegraphics[height=3cm]{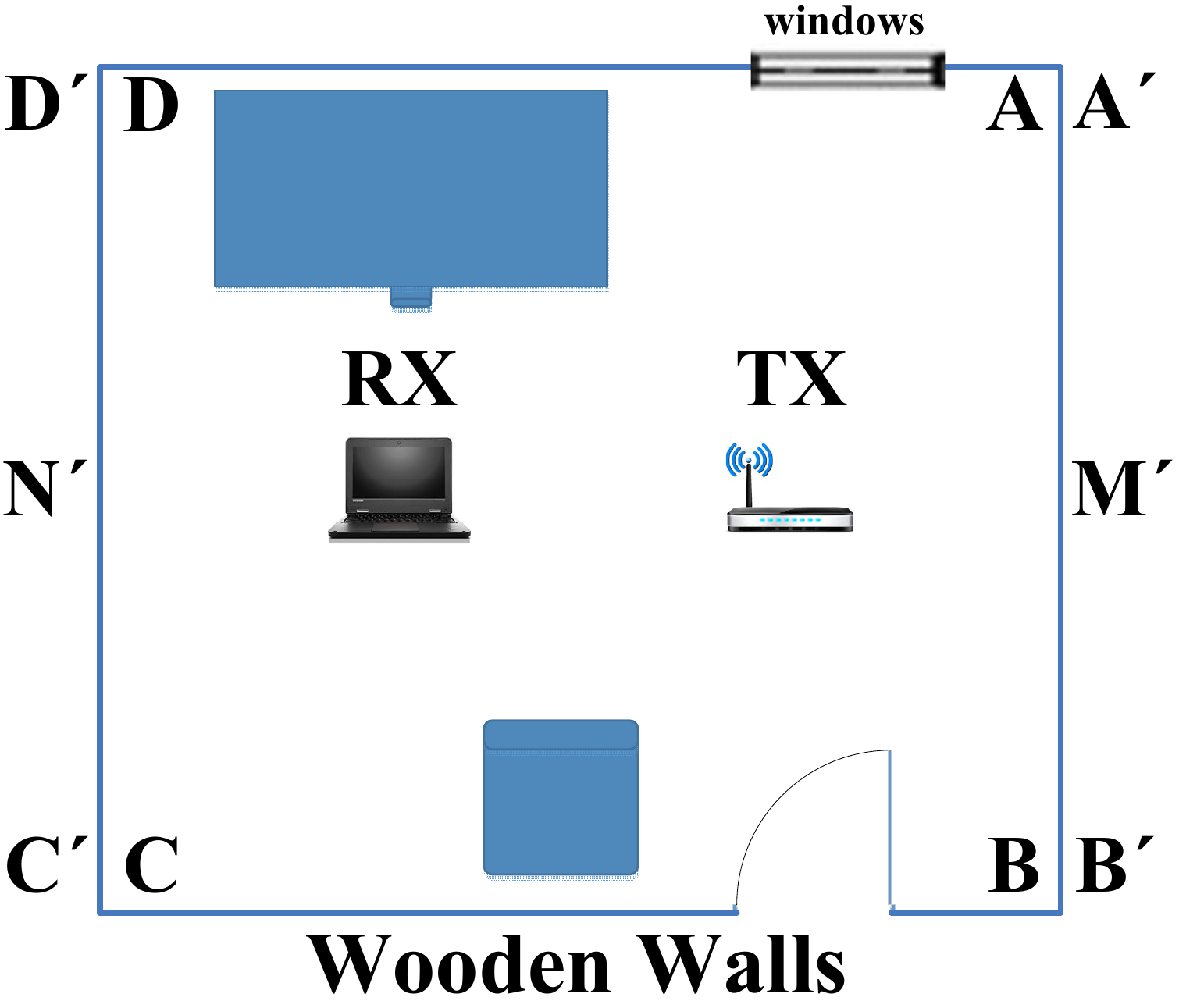}
}
\caption{Square room: two deployment configurations.}
\label{fig:config-squre-room}
\end{figure}

We deploy the \texttt{VSButton} in a wooden square room and evaluate it with two deployment configurations as shown in Figure~\ref{fig:config-squre-room}. In the first configuration, the laptop with an Echo dot (RX) is placed at the center of the room and the WiFi router (TX) is located at the edge. In the second configuration, the RX and the TX are placed between Locations $N'$ and $M'$ to divide the distance into three equidistant portions.
In the experiments, the six participants do the aforementioned three motions (i.e., WAVE-HAND, SIT-DOWN-STAND-UP, and JUMP) at four indoor locations ($A$, $B$, $C$, and $D$) and six outdoor locations ($A'$, $B'$, $C'$, $D'$, $M'$, and $N'$).

\smallskip
\noindent \textbf{Configuration 1.} 
The Mahalanobis distances we measured are summarized in Table~\ref{tbl-max-distance-for-squre-room-config1}. Note that each number of the indoor results is the minimum value of all the numbers measured among the participants, whereas that of the outdoor results is the maximum.
This way can easily show whether indoor and outdoor motions can be clearly differentiated based on the Mahalanobis distances or not.
We observe that all the indoor motions can be differentiated from no-motion cases and outdoor motions at all the locations except Location $M'$. Some indoor motions, such as the WAVE-HAND motion at Location $D$, have smaller distances than the motions, JUMP and SIT-DOWN-STAND-UP, at Location $M'$. The main reason is that that location is very close to where the WiFi router is deployed. As a result, the router shall not be deployed at the location close to the wall next to outdoor space.



\smallskip
\noindent
\textbf{Configuration 2.} 
The results of this configuration are summarized in Table~\ref{tbl-max-distance-for-squre-room-config2}.
We observe that the distance of each indoor motion is higher than the maximum distance (i.e., $0.241$ from JUMP at Location $M'$)of all the outdoor motions.
%
%
It represents that \texttt{VSButton} can activate the Alexa service only due to indoor motions with this configuration.
\subsubsection{A Rectangle Lab Room}

\begin{figure}[t]
\centering
\includegraphics[width=0.8\columnwidth]{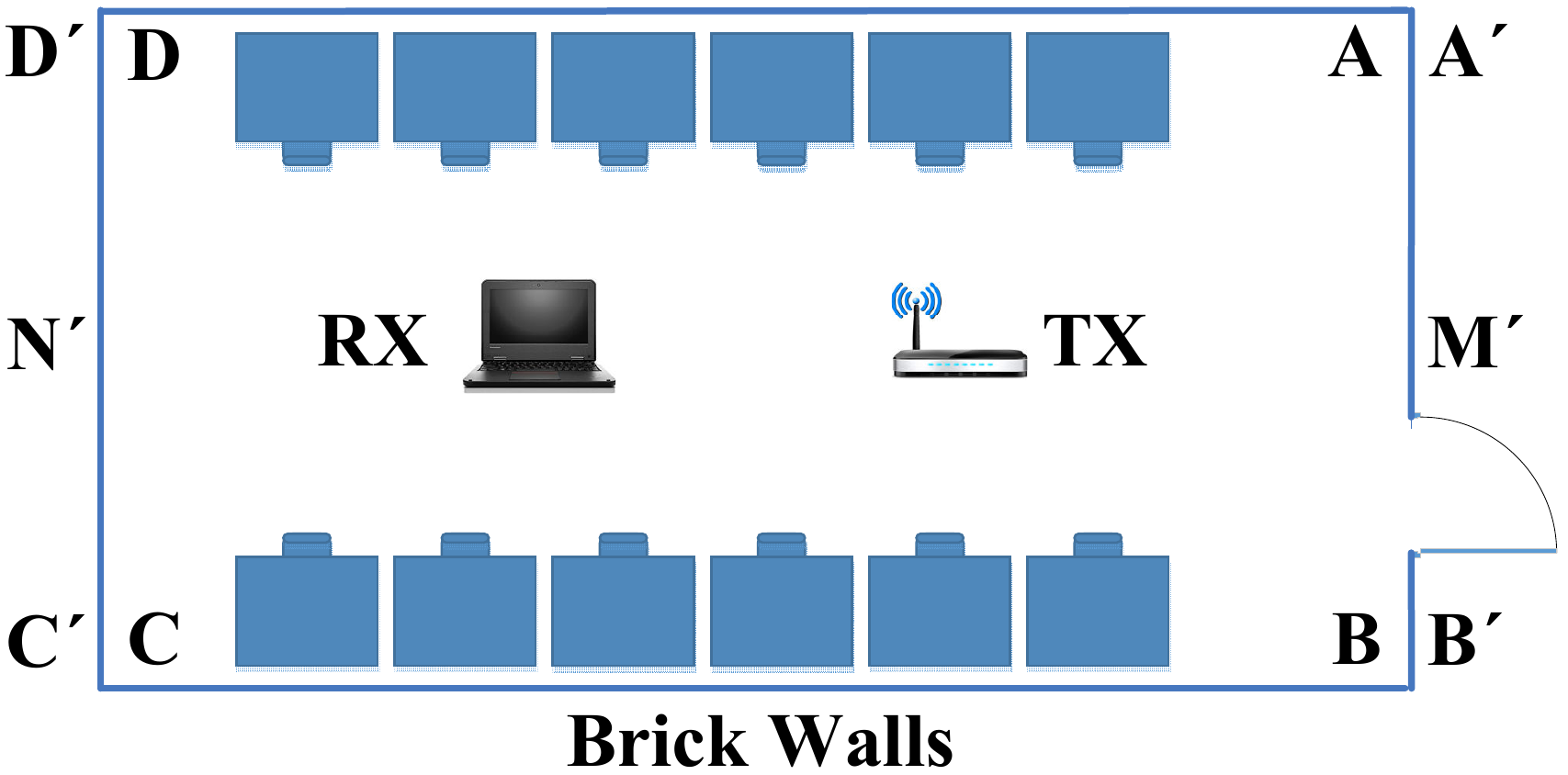}
\caption{Rectangle Room Configuration.}
\label{RR}
\end{figure}

We now evaluate the prototype in a rectangle room with brick walls. The RX and the TX are placed between Locations $N'$ and $M'$ to divide the distance into three equidistant portions, as shown in Figure~\ref{RR}. Table~\ref{tbl-max-distance-for-rectangle-room-config1} summarizes the measurement results.
Note that the distance at each indoor location is the minimum value of all the numbers measured among the participants, whereas that at each outdoor location is the maximum.

\begin{table*}[t]
\centering
\resizebox{0.95\textwidth}{!}{
\normalsize
\renewcommand{\arraystretch}{1}
\begin{tabular}{|l|c|c|c|c|c|c|c|c|c|c|}
\hline
 \textbf{Rectangle Room}      &  \multicolumn{4}{|c|}{\textbf{Indoor locations}} & \multicolumn{6}{|c|}{\textbf{Outdoor locations}}\\
\hline
   locations & $A$  & $B$ & $C$  & $D$ & $A'$ & $B'$ &  $C'$ & $D'$ & $M'$ &  $N'$  \\
\hline
\hline
WAVE-HAND  & \underline{$\mathbf{0.147}$}   & $0.150$ & $0.180$ & $0.183$ & $0.020$ & $0.022$ & $0.025$  & $0.027$ & $0.035$ & $0.030$\\
\hline
SIT-DOWN-STAND-UP   & $0.181$ & $0.184$   & $0.216$ & $0.217$ & $0.024$ & $0.026$ & $0.028$ & $0.029$ & $0.039$ & $0.033$  \\
\hline
JUMP  & $0.254$  & $0.255$ & $0.287$ & $0.288$ & $0.029$ & $0.029$ & $0.032$ & $ 0.033$ & \underline{$\mathbf{0.042}$} & $0.035$ \\
\hline
DO NOTHING  & $0.022$  &  $0.021$ & $0.022$ & $0.027$ &  $0.028$ & $0.026$ & $0.021$ & $0.022$ & $0.020$ & $0.025$ \\
\hline
\end{tabular}
}
\caption{Mahalanobis distance measured  in a rectangle room.}
\label{tbl-max-distance-for-rectangle-room-config1}
\end{table*}
\normalsize

There are two findings. First, the result is similar to that of the square room with the same configuration (i.e., Configuration 2). The minimum Mahalanobis distance (i.e., $0.147$ from WAVE-HAND at Location $A$) among all indoor motions is higher than the maximum distance (i.e., $0.042$ from JUMP at Location $M'$) of all outdoor motions. Their difference is as large as $0.105$.  Second, that difference is higher than that (i.e., $0.071$) observed in the square room with the same configuration. The main reason is that WiFi signals are unable to penetrate the brick walls as easy as wooden walls.

There are two lessons learned. First, the wall materials can influence the performance of \texttt{VSButton}.
The harder the wall materials are, the better performance the \texttt{VSButton} can get.
Our experimental results show that \texttt{VSButton} can be applied to two kinds of wall materials, wood and brick. Second, users should not deploy \texttt{VSButton} at the location which is close to outdoor space.


%
%

\subsubsection{An Apartment with Two-bedroom and One-bathroom}

\begin{figure}[t]
\centering
\includegraphics[width=0.9\columnwidth]{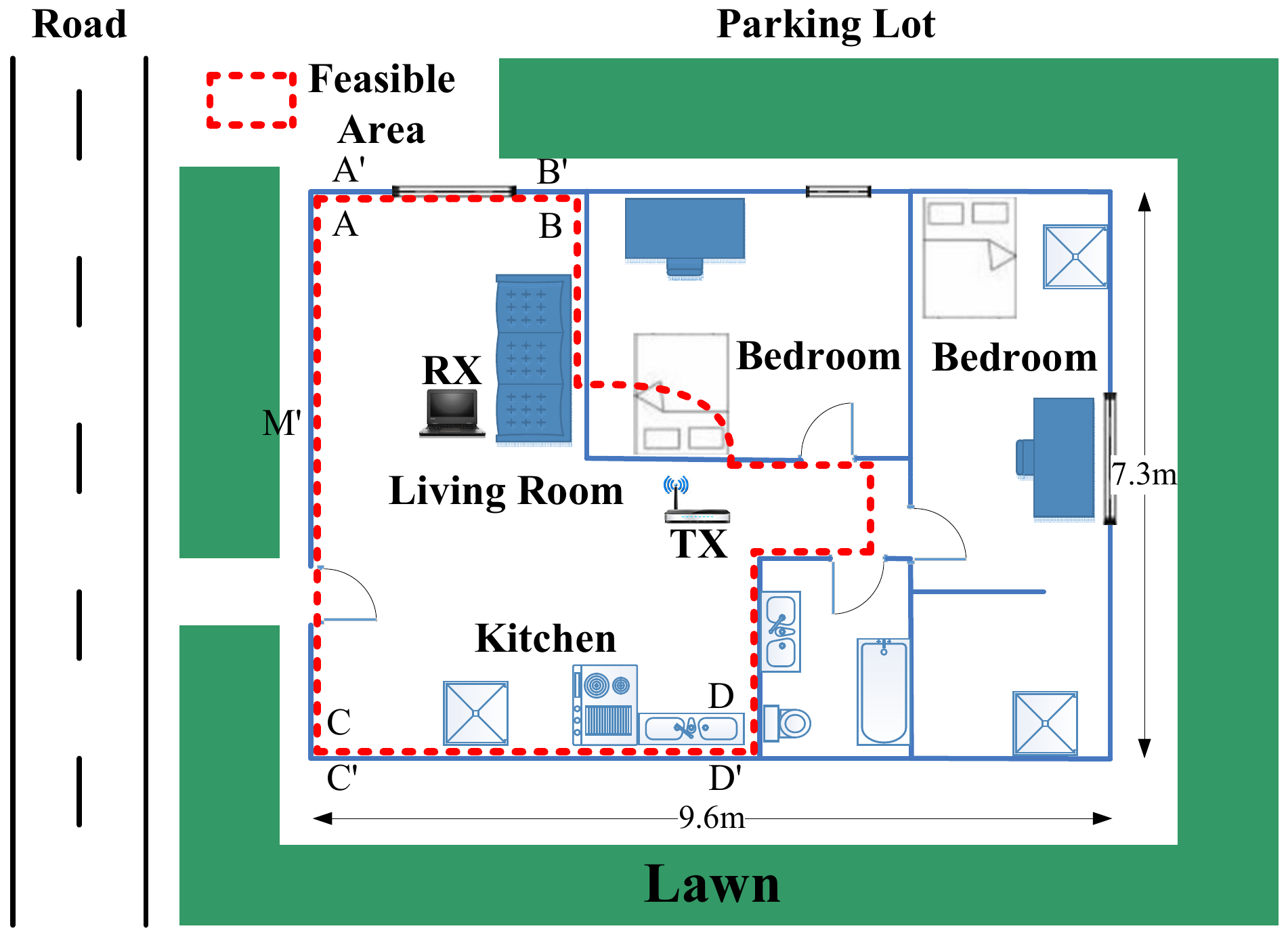}
\caption{\texttt{VSButton} Deployment in an apartment with two bedrooms and one bathroom.}
\label{2b-condo}
\end{figure}

We also evaluate \texttt{VSButton} in a 75$m^2$ apartment with two bedrooms and one bathroom, as shown in Figure~\ref{2b-condo}. We deploy the WiFi router (TX) and the \texttt{VSButton} prototype (RX) at the center of the apartment and the living room, respectively.
In the following, we first do parameter calibration to determine the threshold which differentiates indoor motions from the others, and then examine the result of a 100-minute experiment.

\smallskip
\noindent
\textbf{Parameters Calibration.} Before deploying the \texttt{VSButton} in a real-world scenario, we need to perform parameter calibration to determine a proper threshold $t$ for the outlier motion detection module. It is because the threshold can change with different environments.

The calibration process includes two major steps. 
First, the Alexa owner chooses an indoor location to deploy his/her Alexa device and then determines which area is allowed to enable the device with human motions. At that location, s(he) does the smallest indoor motion (e.g., waving a hand) and collects its minimum Mahalanobis distance value.
Second, the owner finds all the outdoor locations which are not allowed to enable the Alexa device. S(he) does the strongest outdoor motion (e.g., jumping) and collects its maximum Mahalanobis distance value.
We then set the threshold $t$ to be the half of the difference between the above two distance values.
Note that the whole calibration process can be done within 5 minutes and only one-time calibration is needed for the initial deployment of the Alexa devices.
\add{Our solution doesn't require manual re-calibration, but only initial calibration. It is because all the parameters in Table \ref{tble-exp-settings} are fixed and optimized in our design, but only the threshold t and the CSI baselines need to be adapted to environment changes. Specifically, the threshold t is affected only by wall materials, so it requires only initial calibration if the wall is not altered. The CSI baselines, which are used to indicate the conditions of no indoor motions, are automatically, dynamically adapted to different environments (e.g., a new object is deployed nearby).}


The parameter calibration for the apartment is conducted as follows. We perform the WAVE-HAND motion at four indoor locations, $A$, $B$, $C$, and $D$, and the JUMP motion at five outdoor locations, $A'$, $B'$, $C'$, $D'$, and $M'$. We collect the Mahalanobis distance values of difference cases, as shown in Figure~\ref{tbl-max-distance-for-rectangle-room-config1}. The threshold $t$ is set to 0.1, which can differentiate indoor and outdoor locations.

\smallskip
\noindent
\textbf{100-minute Experiment.}
In a 100-minute experiment, we observe that the Alexa device can be activated by the WAVE-HAND motion which happens in the area surrounded by the red dash line (shown in Figure~\ref{2b-condo}). It is not activated by those three outdoor motions made by our participants.
Note that we conduct the parameter calibration process when there is only one user in this apartment. However, in this 100-minute experiment, there are more than one participants in this apartment. Only one participant is allowed to wave his/her hands and the others sit on the sofa, stand in the living room, or stay in the kitchen without making any motions. It shows that the \texttt{VSButton} system is not sensitive to new indoor objects (i.e., participants).



\section{Discussions}
\label{sect:discussions}

We next discuss several remaining issues.

\smallskip
\noindent\textbf{Motions not from Humans.} In our current prototype, we do not consider the motions which are not from humans. For example, the \texttt{VSButton} may activate the Alexa service due to the jump of a pet when the owner is not around. However, it may not threaten the Alexa's security since no voice commands are delivered right after the activation. 

\smallskip
\noindent\textbf{Limited usage locations.} 
\add{\texttt{VSButton} supports Alexa usage only when the command comes from the same room with Alexa. Users usually command the Alexa within the same room due to two reasons. First, they would like to see the light flashing on the Alexa to confirm that it is ready for voice commands. Second, the sounds generated outside the room may not be loud enough to command the Alexa. So, our target scenario, where voice commands are produced within the same room as the Alexa, doesn¡¯t limit the Alexa¡¯s current usage locations.}

\smallskip
\noindent\textbf{Limited Floor Plan.} In this work, we evaluate the \texttt{VSButton} performance in three kinds of floor plans, a square room, a rectangle room, and a 2-bedroom apartment, which are popular in the U.S. For example, ApartmentList.com shows that the number of 1-bedroom and 2-bedroom rents is the median among all the units available in 100 largest U.S. cities~\cite{NationalRent}.
In our future work, we plan to examine more floor plans.

\smallskip
\noindent\textbf{Why not Motion Sensors?}
\texttt{VSButton} is better than motion sensors due to two reasons.
First, most of COTS motion sensors are developed on top of the infrared technology. The detection range and the width angle are approximately 5-8 meters and 45\%-120\%, respectively, if there are no obstacles between moving objects and sensors.
However, \texttt{VSButton} has longer detection range and is not restricted by light-of-sight propagation, which the infrared has. Second, \texttt{VSButton} relies on the WiFi, which already exists in the Alexa devices, but the deployment of the infrared hardware required by motion sensors will introduce extra costs.

\section{Related Work}
\label{sect:related}

\noindent\textbf{WiFi-based Indoor Human Activity/Motion Sensing.} There have been several related works~\cite{Kosba:6199865, Zeng:2014:YAK:2643614.2643620, Pu:2013:WGR:2500423.2500436, Wang:2014:EDL:2639108.2639143} which study how to detect indoor human motions or activities by WiFi technologies. Kosba et al.~\cite{Kosba:6199865} propose to use RSS (received signal strength) to detect human motions by RASID. However, due to the limitation of RSS (i.e., providing coarser granularity of wireless channel information than CSI), RASID is mainly designed to detect relative larger motions, e.g., walking, instead of small motions, e.g., wave a hand. Pu et al.~\cite{Pu:2013:WGR:2500423.2500436} leverages Doppler shift to recognize nine whole-home gestures, requiring specialized receiver that extracts carrier wave features that are not reported in current WiFi systems. However, \texttt{VSButton} is developed on top of off-the-shelf equipments.

Zeng et al.~\cite{ Zeng:2014:YAK:2643614.2643620} adopt CSI-sensing and supervised machine learning techniques to identify four types of indoor motions. Wang et al.~\cite{Wang:2014:EDL:2639108.2639143} leverages CSI to detect fine grained human activities (e.g., walking or cooking). 
\add{CSI-based motion recognition techniques have two limitations to be used in this work. First, they are sensitive to the changes of the environment (e.g., moving the furniture) and sensitive to the location of human. Second, they need laborious pre-training process.
Differ from them, the \texttt{VSButton} just needs one-time set up of a parameter. It detects outlier from the unstable CSI stream and dynamically adapts CSI baselines, which are used to indicate the conditions of no indoor motions, to different environments over time. However, they are absent in the prior studies monitoring CSI values for gesture recognition. They can recognize only different gestures, and be also very sensitive to the environment changes due to its required detection of small CSI variations.}

\smallskip
\del{\noindent\textbf{Software Voice Assistants.}
There are several research works focusing on acoustic attacks towards the software voice assistants on smartphones \cite{vaidya2015cocaine, carlini2016hidden}.
Mangled voice attack is first proposed in \cite{vaidya2015cocaine} and further developed in \cite{carlini2016hidden}. They show that the software voice assistant (Google Voice) on phones can receive voice commands that are unrecognizable to human but interpretable by the voice assistants. The attack works if adversaries (speakers) to be not more than 3.5 meters far away from the phones and the victims do not notice the hearable mingle voice commands. Differ from them, we study home digital voice assistants instead of software voice assistants on smartphones. Their use scenarios and security issues are different. For example, users usually have their smartphones with them, whereas HDVA users leave the home DAV devices at homes.}

\smallskip
\noindent\textbf{Mangled \& Inautdiable Voice Attack.}
\add{Mangled voice attack is first proposed in \cite{vaidya2015cocaine} and further developed in \cite{carlini2016hidden}. They show that the software voice assistant (Google Voice) on phones can receive voice commands that are unrecognizable to human but interpretable by the voice assistants. The attack works if adversaries (speakers) to be not more than 3.5 meters far away from the phones and the victims do not notice the hearable mingle voice commands. Differ from them, we study home digital voice assistants instead of software voice assistants on smartphones. Their use scenarios and security issues are different. For example, users usually have their smartphones with them, whereas HDVA users leave the home DAV devices at homes. The inaudible voice commands attacks towards Amazon Echo is provided in \cite{Song2017}. The attacks require two strong perquisites, which may be barely satisfied in practice, so they are not our focus. First, the inaudible voice commands need to be generated from a customized ultrasound microphone. Second, the customized microphone requires being deployed next to the Alexa device within 2 meters.
}

\smallskip
\noindent\textbf{Voice Authentication for Digital Voice Assistants.}
The Biometric-based voice authentication~\cite{das2008multilingual, kunz2011continuous, baloul2012challenge} have been proposed to protect HDVA devices against acoustic attacks. However, they still have some challenges to address. For example, human's voice will change with age, illness and tiredness. This approach may require a re-training process periodically. Moreover, human's voice can be duplicated by computers by deep learning in a minute~\cite{duplicateHumanVoice2017}.
In addition to biometric-based voice authentication mechanism,~\cite{feng2017continuous} proposed that a DVA user wears a voice-enabled device that touches his/her skin. When the user speaks voice commands, the wearable sensor collects the body-surface vibrations signal and continuously match it with the voice commands received by the DVA. This approach does require the extra wearable devices but also hurt the user convenience.
Differ from them, \texttt{VSButton} well address all of the above challenges while preserving the user convenience.
\texttt{VSButton} authenticates users if they can prove that they are in the indoor space where the HDVA device is deployed. As a result, adversaries cannot launch remote acoustic attacks. \texttt{VSButton} does not suffer from variant human voice, extra deployment cost or the scarifying of user inconvenience (e.g., put wearable devices).

\section{Conclusions}
\label{sect:concl}

HDVAs enable users to control smart devices and get living assistance (e.g., online shopping) using voice commands.
However, this convenience may come with security threats. 
In this work, we identify several security vulnerabilities by considering Amazon Alexa as a case study.
They span all the involved parties, which include Alexa devices, Alexa service providers, and third party Alexa service developers. Surprisingly, the Alexa service relies on only a weak single-factor authentication, which can be easily broken.
Without a physical presence based access control, the Alexa device accepts voice commands even when no persons are nearby.
This security threat can easily propagate to the third party voice services, which assume all the voice commands from the Alexa are benign. We then devise two proof-of-concept attacks to show that such security threat can cause Alexa users to experience a large financial loss.

To secure the HDVA service, we seek to propose an additional factor authentication, physical presence. An HDVA device can accept voice commands only when any person is physically present nearby it. We thus design a solution called virtual security button (\texttt{VSButton}) to do the physical presence detection. We prototype and evaluate it with an Alexa device.
Our experimental results show that \texttt{VSButton} can do accurate detection in both the laboratory and real-world home settings. We hope our initial efforts can stimulate further research on this HDVA security topic.

\bibliographystyle{IEEEtran}
\bibliography{Alexa}

\begin{thebibliography}{10}
\providecommand{\url}[1]{#1}
\csname url@samestyle\endcsname
\providecommand{\newblock}{\relax}
\providecommand{\bibinfo}[2]{#2}
\providecommand{\BIBentrySTDinterwordspacing}{\spaceskip=0pt\relax}
\providecommand{\BIBentryALTinterwordstretchfactor}{4}
\providecommand{\BIBentryALTinterwordspacing}{\spaceskip=\fontdimen2\font plus
\BIBentryALTinterwordstretchfactor\fontdimen3\font minus
  \fontdimen4\font\relax}
\providecommand{\BIBforeignlanguage}[2]{{%
\expandafter\ifx\csname l@#1\endcsname\relax
\typeout{** WARNING: IEEEtran.bst: No hyphenation pattern has been}%
\typeout{** loaded for the language `#1'. Using the pattern for}%
\typeout{** the default language instead.}%
\else
\language=\csname l@#1\endcsname
\fi
#2}}
\providecommand{\BIBdecl}{\relax}
\BIBdecl

\bibitem{StrategyDVA2017}
D.~Watkins, ``Strategy analytics: Amazon, google to ship nearly 3 million
  digital voice assistant devices in 2017,''
  https://www.strategyanalytics.com/strategy-analytics/news/strategy-analytics-press-releases/strategy-analytics-press-release/2016/10/05/strategy-analytics-amazon-google-to-ship-nearly-3-million-digital-voice-assistant-devices-in-2017\#.WQtiXeXyuUk,
  2016.

\bibitem{Sales}
``Amazon: Alexa devices were best-selling products from any manufacturer,''
  \url{https://marketingland.com/amazon-alexa-devices-best-selling-products-manufacturer-199446
  }, 2016.

\bibitem{10000Skills}
``Amazon alexa hits 10,000 skills,''
  \url{https://www.wired.com/2017/02/amazon-alexa-hits-10000-skills-plenty-room-grow/
  }, 2017.

\bibitem{feng2017continuous}
H.~Feng, K.~Fawaz, and K.~G. Shin, ``Continuous authentication for voice
  assistants,'' \emph{arXiv preprint arXiv:1701.04507}, 2017.

\bibitem{lei2018insecurity}
X.~Lei, G.-H. Tu, A.~X. Liu, C.-Y. Li, and T.~Xie, ``The insecurity of home
  digital voice assistants-vulnerabilities, attacks and countermeasures,'' in
  \emph{CNS}, 2018, pp. 1--9.

\bibitem{HackChromeCast}
D.~Petro, ``Rickrolling your neighbors with google chromecast,'' in \emph{Black
  Hat}, 2014.

\bibitem{HackAnswerMachine}
``Hacking answering machines,''
  \url{https://www.youtube.com/watch?v=bq6aV0Cxhl0}, 2012.

\bibitem{HackBluetoothSpeaker}
``How to prevent unauthorized access to bluetooth speakers?''
  \url{https://superuser.com/questions/548592/how-to-prevent-unauthorized-access-to-bluetooth-speakers},
  2013.

\bibitem{Buletooth}
``Connect your echo device to bluetooth speakers,''
  \url{https://www.amazon.com/gp/help/customer/display.html?nodeId=202011820},
  2017.

\bibitem{RecoverWiFiRouter}
``How to recover wifi password in android without root?''
  \url{http://www.viralhax.com/recover-wifi-password-android/}, 2017.

\bibitem{DefaultPassofWirelessRouter}
``Default username and password of wireless routers,''
  \url{http://www.routerpasswords.com/}, 2017.

\bibitem{Garageio}
``Work with garageio,'' \url{https://garageio.com/workswith/echo}, 2017.

\bibitem{jolliffe2002principal}
I.~Jolliffe, \emph{Principal component analysis}.\hskip 1em plus 0.5em minus
  0.4em\relax Wiley Online Library, 2002.

\bibitem{arce2005nonlinear}
G.~R. Arce, \emph{Nonlinear signal processing: a statistical approach}.\hskip
  1em plus 0.5em minus 0.4em\relax John Wiley \& Sons, 2005.

\bibitem{holt2004forecasting}
C.~C. Holt, ``Forecasting seasonals and trends by exponentially weighted moving
  averages,'' \emph{International journal of forecasting}, vol.~20, no.~1, pp.
  5--10, 2004.

\bibitem{butterworth1930theory}
S.~Butterworth, ``On the theory of filter amplifiers,'' \emph{Wireless
  Engineer}, vol.~7, no.~6, pp. 536--541, 1930.

\bibitem{abdi2010principal}
H.~Abdi and L.~J. Williams, ``Principal component analysis,'' \emph{Wiley
  interdisciplinary reviews: computational statistics}, 2010.

\bibitem{moshtaghi2011incremental}
M.~Moshtaghi, C.~Leckie, S.~Karunasekera, J.~C. Bezdek, S.~Rajasegarar, and
  M.~Palaniswami, ``Incremental elliptical boundary estimation for anomaly
  detection in wireless sensor networks,'' in \emph{ICDM}, 2011.

\bibitem{IEEE802}
``Ieee std. 802.11n-2009: Enhancements for higher throughput,''
  \url{http://www.ieee802.org}, 2009.

\bibitem{AlexaWifiSpec}
``Connect echo dot to wi-fi,''
  \url{https://www.amazon.com/gp/help/customer/display.html?nodeId=202011800},
  2017.

\bibitem{halperin2011tool}
D.~Halperin, W.~Hu, A.~Sheth, and D.~Wetherall, ``Tool release: Gathering
  802.11 n traces with channel state information,'' in \emph{SIGCOMM}, 2011.

\bibitem{MicroBotPush}
``Microbot push,'' \url{https://prota.info/}, 2017.

\bibitem{NationalRent}
``Apartment list national rent report,''
  \url{https://www.apartmentlist.com/rentonomics/national-rent-data/}, 2017.

\bibitem{Kosba:6199865}
A.~E. Kosba, A.~Saeed, and M.~Youssef, ``Rasid: A robust wlan device-free
  passive motion detection system,'' in \emph{PerCom}, 2012.

\bibitem{Zeng:2014:YAK:2643614.2643620}
Y.~Zeng, P.~H. Pathak, C.~Xu, and P.~Mohapatra, ``Your ap knows how you move:
  Fine-grained device motion recognition through wifi,'' in \emph{HotWireless},
  2014.

\bibitem{Pu:2013:WGR:2500423.2500436}
Q.~Pu, S.~Gupta, S.~Gollakota, and S.~Patel, ``Whole-home gesture recognition
  using wireless signals,'' in \emph{Mobicom}, 2013.

\bibitem{Wang:2014:EDL:2639108.2639143}
Y.~Wang, J.~Liu, Y.~Chen, M.~Gruteser, J.~Yang, and H.~Liu, ``E-eyes:
  device-free location-oriented activity identification using fine-grained wifi
  signatures,'' in \emph{MobiCom}, 2014.

\bibitem{vaidya2015cocaine}
T.~Vaidya, Y.~Zhang, M.~Sherr, and C.~Shields, ``Cocaine noodles: exploiting
  the gap between human and machine speech recognition,'' in \emph{WOOT}, 2015.

\bibitem{carlini2016hidden}
N.~Carlini, P.~Mishra, T.~Vaidya, Y.~Zhang, M.~Sherr, C.~Shields, D.~Wagner,
  and W.~Zhou, ``Hidden voice commands,'' in \emph{USENIX Security}, 2016.

\bibitem{Song2017}
L.~Song and P.~Mittal, ``Inaudible voice commands,''
  \url{https://arxiv.org/pdf/1708.07238.pdf}, 2017.

\bibitem{das2008multilingual}
A.~Das, O.~K. Manyam, M.~Tapaswi, and V.~Taranalli, ``Multilingual
  spoken-password based user authentication in emerging economies using
  cellular phone networks,'' in \emph{SLT}, 2008.

\bibitem{kunz2011continuous}
M.~Kunz, K.~Kasper, H.~Reininger, M.~M{\"o}bius, and J.~Ohms, ``Continuous
  speaker verification in realtime.'' in \emph{BIOSIG}, 2011.

\bibitem{baloul2012challenge}
M.~Baloul, E.~Cherrier, and C.~Rosenberger, ``Challenge-based speaker
  recognition for mobile authentication,'' in \emph{BIOSIG}, 2012.

\bibitem{duplicateHumanVoice2017}
L.~Dormehl, ``Ai can now duplicate anyone's voice based on just one minute of
  training.''
  \url{https://www.digitaltrends.com/cool-tech/ai-lyrebird-duplicate-anyones-voice/},
  2017.

\end{thebibliography}

\end{document}